\documentclass[a4paper,aps,prx,floatfix,twocolumn, superscriptaddress,nobibnotes,longbibliography]{revtex4-2}

\usepackage[pdftex]{graphicx}
\setkeys{Gin}{width=0.9\columnwidth}

\usepackage{times}
\usepackage{verbatim}
\usepackage{bbold}
\usepackage{bbm}
\usepackage{lipsum}

\usepackage{amsmath}
\usepackage{amsfonts}
\usepackage{color}
\usepackage[english]{babel}
\usepackage{microtype}
\usepackage{comment}
\usepackage[normalem]{ulem}
\usepackage{float}

\usepackage{soul,xcolor}
\setstcolor{red}

\usepackage{relsize}

\usepackage[hidelinks,unicode=true]{hyperref}
\hypersetup{
colorlinks=true,
linkcolor=blue,
urlcolor=blue,
citecolor=blue,
pdfhighlight=/N}

\usepackage{comment}

\renewcommand{\sout}[1]{\unskip}

\newcommand{\add}[1]{{#1}}
\newcommand{\delete}[1]{\sout{#1}}

\usepackage{relsize}
\newcommand{\MS}[1]{{\color{cyan} (MS: {#1})}}

\newcommand{\av}[1]{\langle #1 \rangle}

\newcommand{\im}[1]{\mathrm{Im}\left\{ {#1} \right\}}
\newcommand{\re}[1]{\mathrm{Re}\left\{ {#1} \right\}}
\DeclareMathOperator\arctanh{arctanh}

\begin{document}

\title{Social Network Heterogeneity Promotes Depolarization of Multidimensional Correlated Opinions}

\author{Jaume Ojer}

\affiliation{Departament de F\'isica, Universitat Polit\`ecnica de Catalunya, Campus Nord, 08034 Barcelona, Spain}

\author{Michele Starnini}
\email{Corresponding author: michele.starnini@gmail.com}
\affiliation{Departament de F\'isica, Universitat Polit\`ecnica de Catalunya, Campus Nord, 08034 Barcelona, Spain}

\affiliation{CENTAI Institute, 10138 Turin, Italy}

\author{Romualdo Pastor-Satorras}
\email{Corresponding author: romualdo.pastor@upc.edu}
\affiliation{Departament de F\'isica, Universitat Polit\`ecnica de Catalunya, Campus Nord, 08034 Barcelona, Spain}

\date{\today}

\begin{abstract}
Understanding the mechanisms to mitigate opinion polarization in our society is crucial to minimizing social division and ultimately strengthening democracy. 
Due to the challenge of collecting long-term reliable empirical data, researchers have been mostly focused on a theoretical understanding of the process of opinion depolarization. 
To this aim, realistic yet simple models prove valuable, especially when multiple topics are discussed at the same time, which may result in entangled opinion dynamics. 
In this paper, we propose the multidimensional social compass model, based on 
two competing key ingredients: DeGroot learning, driven by the social influence exerted across multiple topics, 
and the preference of individuals to maintain
their initial opinions. 
The interplay between these two mechanisms triggers a phase transition from polarization to consensus, determined by a threshold value of social influence. 
We analytically study the nature of the depolarization transition and its threshold depending on the number of topics discussed, the possible correlations between initial opinions, the topology of the underlying social networks, and the correlations between the initial opinion distribution and the network's structure. 
Theoretical predictions are validated by running numerical simulations on both synthetic and real social networks. 
We rely on several simplifying assumptions to explore different scenarios, such as a mean-field approximation for high dimension, or orthogonal initial orientations. 
We uncover an upper critical dimension ($D_c = 5$ topics) for uncorrelated initial opinions, distinguishing between discontinuous and continuous phase transitions. 
For the simplest $D = 2$ case and correlated initial opinions, we found that the depolarization threshold can vanish if the underlying connectivity is heterogeneous, 
as predicted by perturbation theory. 
Such an effect is due to the presence of hubs, which promote consensus in the population. 
We test this hypothesis by designing a rewiring
algorithm that increases the structural heterogeneity of the underlying network, showing that the depolarization threshold decreases. 
Finally, we demonstrate that if hubs share the same initial opinion, the depolarization dynamics is significantly hindered. 
Our findings contribute to understanding the mechanisms to mitigate polarization in real-world scenarios, suggesting which settings can promote the depolarization process. 
The presence of very popular individuals on online social networks and the alignment of their opinions, in particular, may play a pivotal role in the multidimensional depolarization dynamics.

\end{abstract}

\maketitle

\section{Introduction}

The emergence of opinion polarization, which usually refers to the presence of two groups in a population holding opposite and potentially extreme opinions~\cite{mccoy_polarization_2018}, can have a negative impact on society, providing a breeding ground for the dissemination of harmful misinformation~\cite{vicario_polarization_2019}. 
Consequently, researchers from distinct scientific disciplines have embarked on studying different mechanisms that could potentially reduce opinion polarization, a process referred to as \emph{depolarization} in the literature~\cite{vinokur1978depolarization, miller_attitude_1993}. 
Given the limited availability of reliable data on opinions spanning an extended period, most studies have concentrated on a theoretical understanding of opinion depolarization~\cite{matakos_measuring_2017, musco_minimizing_2018, balietti_reducing_2021}. 
Several modeling efforts have explored various mechanisms underlying the opinion formation process, such as homophily~\cite{axelrod_dissemination_1997, dandekar_biased_2013}, bounded confidence~\cite{deffuant_mixing_2000, hegselmann_opinion_2002}, or opinion rejection~\cite{martins_mass_2010, crawford_opposites_2013}.

While earlier studies typically focused on a single topic~\cite{sznajd-weron_opinion_2000, lopez-pintado_social_2008}, researchers recently have proposed novel modeling frameworks to address the case of multidimensional opinions~\cite{schweighofer_agent-based_2020, chen_axioms_2021, zhou_multidimensional_2022}, where opinions can vary across different topics. 
Notably, when multiple topics are considered, correlations between opinions may arise~\cite{baldassarri_partisans_2008, benoit_dimensionality_2012, dellaposta_why_2015, falck_measuring_2020}. 
These correlations emerge when topics are interdependent, i.e., when the positions with respect to certain issues are aligned and thus the resulting opinion dynamics regarding these topics are entangled~\cite{baumann_emergence_2021}. 
As a consequence, individuals are much more likely to have a certain combination of opinions than others, a state defined as an \emph{ideology} in political science. 
For instance, racism and the stereotype of blacks as violent individuals may be related to gun ownership and opposition to gun control policies in US whites~\cite{obrien_racism_2013}. 
Importantly, issue alignment can increase political polarization via partisan sorting, i.e., voters sort themselves into parties that match their ideology~\cite{mason_i_2015}. 

Within a statistical physics modeling framework, the process of opinion depolarization can result in a phase transition from an initially polarized opinion state to a consensus~\cite{currin_depolarization_2022, pal_depolarization_2023, sobkowicz_social_2023}. 
The presence of correlations among opinions in multidimensional topic models may play a pivotal role in the phase transition from polarization to consensus.
For instance, very recently, it has been shown that the nature of the depolarization transition can change from first to second order if topics are interdependent, triggering an explosive depolarization in the independent topic case~\cite{ojer_modeling_2023} . 
The basic mechanism responsible for a depolarization transition is typically related to the social influence exerted by individuals on their peers. 
Such a social influence mechanism is usually assumed to be mediated by social interactions, by which individuals exchange information and form their opinions~\cite{Jackson2010}. 
In opinion dynamics models, this assumption usually translates into agents being placed on social networks, whose nontrivial connectivity emerges from the self-organization of individuals. 
Crucially, the network topology has been shown to play a decisive role in many networked dynamical processes~\cite{10.5555/1521587}, e.g., spreading dynamics on heterogeneous networks results in a vanishing epidemic threshold~\cite{boguna_absence_2003}. 
However, the effects of the underlying network topology have not been explored yet within the context of multidimensional opinion dynamics models. 
Understanding the impact of social heterogeneity on the depolarization process is crucial to determine under which conditions the population can reduce opinion polarization more effectively.

In this paper, we address this important open question by proposing and analytically studying a general multidimensional opinion depolarization model, the \emph{multidimensional social compass model}. 
The model is based on two key assumptions: i) Individuals 
exert social influence through a social network; ii) They are stubborn to their preferred (initial) opinion. 
This general framework has been formalized for scalar (one-dimensional) opinions by the Friedkin-Johnsen model~\cite{friedkin_social_1990}. 
Here we consider a $D$-dimensional version with interdependent topics, where interaction terms of social influence and stubbornness are formulated as simple differences between opinion 
vectors, modulated by a single parameter. 
Interestingly, by imposing a constant modulus of the opinion vector (to avoid divergence to infinity), the model undergoes a phase transition from polarization to consensus, a behavior also found in models of collective motion~\cite{Cavagna.2018}. 
We consider initially fully polarized opinions, distinguishing between the cases of correlated and uncorrelated ones.

First, 
we obtain analytical expressions for the threshold value of the social influence needed to trigger the depolarization of the population, in the mean-field approximation. 
We discover an upper critical dimension $D_c = 5$ for the uncorrelated case, separating a discontinuous phase transition ($D < D_c$) from a continuous one ($D \ge D_c$), while the correlated case always shows a continuous phase transition with the same threshold for any $D$. 
In light of this crucial insight, we focus next on the $D = 2$ case for correlated opinions and explore the effects of nontrivial underlying connectivity, by exploring two different theoretical approaches to investigate the conditions for the emergence of consensus. 
We show that the threshold value for depolarization is entirely determined by the network topology. 
Such a depolarization threshold is very small for heterogeneous (thus realistic) networks, where the social influence needed to trigger depolarization vanishes in the limit of large networks. 
Therefore, heterogeneous contacts can be beneficial to counterbalance social polarization, with the highly connected nodes fostering consensus among the agents. 
We validate our theoretical findings by running extensive simulations of the depolarization model on both synthetic and empirical social networks, confirming that the initial polarized state is very susceptible to disappear even when social influence among individuals is very weak. 
We further check this hypothesis by proposing and testing an algorithm that rewires the underlying social networks. 
We show that the depolarization threshold decreases as the network's heterogeneity increases: the presence of hubs is key to fostering consensus. 
Finally, we explore the effects of correlations between the network topology and initial opinions. 
We show that, for both synthetic and empirical networks, when hubs share the same initial opinions, a consensus is much harder to achieve.

The paper is organized as follows. 
In Section~\ref{sec:d-dim} we define the multidimensional social compass model and find a mean-field solution for the depolarization threshold in both correlated and uncorrelated initial opinions. 
We also establish a direct connection with the $D=2$ case. 
In Section~\ref{sec:vanish}, we demonstrate a vanishing threshold of the depolarization transition triggered by correlated opinions in a nontrivial connectivity for $D = 2$. We contrast a heterogeneous mean-field approximation with a perturbation theory, showing that only the latter can correctly capture the depolarization threshold in the thermodynamic limit. 
In Section~\ref{sec:hubs}, we test the effect of the network's heterogeneity by proposing a rewiring algorithm, showing the role of hubs in promoting consensus. 
In Section~\ref{sec:patrizi}, we show the effects of correlations between degree and initial opinions, for both synthetic and empirical networks. Finally, we discuss our results and explore possible perspectives in Section~\ref{sec:discussion}.

\section{The multidimensional social compass model}
\label{sec:d-dim}

\subsection{Model definition}

We start by considering 
$N$ individuals expressing opinions regarding $D$ topics. 
Each individual $i$ is characterized by an opinion vector of components $\vec{x}_i = ( x_i^{(1)}, x_i^{(2)}, \dots, x_i^{(D-1)}, x_i^{(D)} )$, where the component $x_i^{\left( \tau \right)} \in \left( -\infty, +\infty \right)$ denotes the opinion towards topic $\tau$. 
The opinion vector $\vec{x}_i$ thus represents the position of an agent $i$ in the $D$-dimensional \textit{topic space}. 
It can be written as $\vec{x}_i = \sum_{\tau=1}^D x_i^{\left( \tau \right)} \hat{e}_\tau$, where the vectors $\{ \hat{e}_\tau : 1 \le \tau \le D \}$ define the standard basis of the topic space $\mathbb{R}^D$. 
The opinion vectors of agents evolve in time, i.e., $\vec{x}_i = \vec{x}_i(t)$, where we will omit the dependence on $t$ in the following for brevity. 
Individuals are characterized by their preferred or initial vector state $\vec{n}_i = \vec{x}_i(0)$, which indicates their preferred opinion in the topic space.

At a general level, the dynamics of an individual's opinion is determined by two competing factors. 
On the one hand, individuals are influenced by the opinions of their peers, with whom they interact and exchange information, contributing to the opinion formation process. 
This mechanism, known as \emph{social influence} or social learning, is usually mediated by a social network that determines the weight that an agent puts on another agent's opinion, as in DeGroot's model~\cite{degroot_reaching_1974}. 
In the absence of other factors, DeGroot learning converges to consensus for strongly connected and aperiodic underlying social networks~\cite{Golub}. 
On the other hand, individuals may have a preferred opinion (sometimes referred to as prejudice) and a tendency to maintain it, i.e., they show a certain degree of \emph{stubbornness}. 
When agents are maximally stubborn, i.e., they never change their opinion, they are sometimes referred to as zealots~\cite{PhysRevE.91.022811}. 
The competition between social influence and stubbornness has been formalized by the Friedkin-Johnsen model~\cite{friedkin_social_1990, zhou_multidimensional_2022}. 
Establishing the conditions for the convergence to consensus in this more complex and realistic setting is more challenging~\cite{parsegov_novel_2017}.

We formalize this general framework into what we call the \emph{multidimensional social compass model}, defined by the set of differential equations~\cite{Cavagna.2018}
\add{\begin{equation}
  \frac{d \vec{x}_i}{d t} = \lambda \left( \sum_{j=1}^N a_{ij} \vec{x}_j - \vec{x}_i \right) + \eta \left( \vec{n}_i
  - \vec{x}_i \right),
  \label{eq:model_D}
\end{equation}}
which fully determine the temporal evolution of the opinion vector $\vec{x}_i(t)$. 
The first term on the right-hand side accounts for the social influence, quantified by the coupling constant $\lambda$. 
Social influence is mediated by a social network with adjacency matrix $a_{ij}$~\cite{Newman10}, taking the value $a_{ij} = 1$ when agents $i$ and $j$ are socially connected and $a_{ij} = 0$ otherwise. 
According to this term, an individual tends to align their opinion $\vec{x}_i$ with the \delete{average opinion} \add{sum of the opinions} of their nearest neighbors $\vec{x}_j$. 
\add{Similarly to social contagion, the probability that an individual adopts a different opinion increases monotonically with the number of neighbors holding that opinion~\cite{hodas_simple_2014}.} 
The second term models the tendency of an individual to remain close to their preferred opinion $\vec{n}_i$, quantified by the parameter $\eta$. 
Parameters $\lambda$ and $\eta$ mediate the interplay between social influence and stubbornness, interpolating between the two extreme cases of completely open-minded ($\eta = 0$) or maximally stubborn ($\lambda = 0$) individuals. 
Since we are not interested in these extreme cases, we fix $\eta = 1$, without loss of generality, by \delete{setting the time unit} \add{re-scaling the time}. 
Therefore, the multidimensional social compass model relies on only two assumptions: i) individuals exert social influence through a social network, and ii) they are stubborn to their preferred (initial) opinion. 
We stress that we choose the simplest linear functional form for the two interaction terms, formulated as differences between opinion vectors.


While simple, the linear Eq.~\eqref{eq:model_D} suffers from the drawback that, for large values of $\lambda$, the modulus of the opinion vector $\vec{x}_i$ grows without bounds. 
One can avoid this by imposing $\vec{x}_i$ to be a unit vector, that is $\vec{x}_i(t) \equiv \hat{x}_i(t)$ with $\left| \hat{x}_i \right| = 1$. 
This can be achieved by adding a Lagrange multiplier to Eq.~\eqref{eq:model_D}, namely~\cite{Cavagna.2018} 
\add{\begin{equation}
  \frac{d \hat{x}_i}{d t} =  \lambda \left( \sum_{j=1}^N a_{ij} \hat{x}_j - \hat{x}_i \right) + \vec{n}_i
  - \hat{x}_i + \Lambda_i \hat{x}_i,
  \label{eq:model_Lagrange}
\end{equation}}
The Lagrange multiplier $\Lambda_i$ is fixed by imposing $\hat{x}_i^2 = 1$, or alternatively $\hat{x}_i \cdot \frac{d \hat{x}_i}{dt} = 0$, leading to
\add{\begin{equation}
    \hat{x}_i \cdot \frac{d \hat{x}_i}{dt} = \lambda \sum_{j=1}^N a_{ij} \hat{x}_j \cdot \hat{x}_i - \lambda + \vec{n}_i \cdot \hat{x}_i - 1 + \Lambda_i = 0.
    \label{eq:Lagrange}
\end{equation}}
We then find
\add{\begin{equation}
    \Lambda_i = 1 - \vec{n}_i \cdot \hat{x}_i - \lambda \left( \sum_{j=1}^N a_{ij} \hat{x}_j \cdot \hat{x}_i - 1 \right).
    \label{eq:Lagrange_multiplier}
\end{equation}}
Inserting it into Eq.~\eqref{eq:model_Lagrange} we obtain
\begin{equation}
  \frac{d \hat{x}_i}{d t} = \lambda \sum_j a_{ij} \left[ \hat{x}_j - (\hat{x}_j \cdot \hat{x}_i) \, \hat{x}_i \right] + \vec{n}_i - (\vec{n}_i \cdot \hat{x}_i) \, \hat{x}_i.
  \label{eq:model_final}
\end{equation}

If we write the preferred opinion as $\vec{n}_i = \rho_i \hat{n}_i$, with $\rho_i$ its modulus and $\hat{n}_i$ its orientation, from Eq.~\eqref{eq:model_final} we obtain
\begin{equation}
  \frac{d \hat{x}_i}{d t}
  =  \lambda \sum_j a_{ij} \left[ \hat{x}_j - \cos(\Phi_{ij}) \, \hat{x}_i \right] + 
  \rho_i \left[ \hat{n}_i - \cos(\Theta_i) \, \hat{x}_i \right],
  \label{eq:model_orientation}
\end{equation}
where we indicated by $\cos(\Phi_{ij})$ the cosine similarity between the orientations of individuals $i$ and $j$, and by $\cos(\Theta_i)$ the cosine similarity between the preferred orientation $\hat{n}_i$ and orientation $\hat{x}_i$ of individual $i$. 
According to Eq.~\eqref{eq:model_orientation}, the orientation of an individual $i$, $\hat{x}_i$, evolves depending on the similarity with the orientations of their neighbors $\hat{x}_j$, and the similarity with their initial and preferred orientation $\hat{n}_i$. 
Interestingly, the strength of the latter coupling is determined by 
the modulus $\rho_i$ of the preferred orientation, representing the \textit{conviction} of the individual $i$ to their initial opinion. 
This interpretation is in line with previous literature related to the effects of the conviction on the opinion formation process~\cite{crokidakis_role_2012, martins_building_2013, burghardt_competing_2016}. 

A vector order parameter for the social compass model can be defined as
\begin{equation}
  \vec{\phi}(t) = \frac{1}{N} \sum_{i=1}^N \hat{x}_i(t),
  \label{eq:vector_order}
\end{equation}
representing the average orientation among the population. 
From here, a scalar order parameter is given by the modulus of the vector order parameter~\cite{Vicsek.1995}
\begin{equation}
  r = | \vec{\phi} | = \frac{1}{N} \left| \sum_i \hat{x}_i \right|.
  \label{eq:scalar_order}
\end{equation}
In the absence of social interactions ($\lambda = 0$), the first term of Eq.~\eqref{eq:model_orientation} accounting for social influence vanishes, leading to the steady state $\hat{x}_i = \hat{n}_i$. Every agent is then aligned with their initial preferred orientation, which is assumed to be a polarized state with a scalar order parameter $r \left( \lambda = 0 \right) = 0$. 
Since the initial polarization is
\begin{equation}
  \vec{\phi}(0) = \frac{1}{N} \sum_i \hat{x}_i(0) = \frac{1}{N} \sum_i \hat{n}_i,
  \label{eq:polarization0}
\end{equation}
an initial state fully polarized is given by preferred orientations such that
\begin{equation}
  \sum_i \hat{n}_i = \vec{0}.
  \label{eq:fully_polarized}
\end{equation}
For large $\lambda$, on the other hand, social influence prevails over stubbornness and a depolarized (consensus) state emerges, with all the agents having the same average orientation $\hat{x}_i \simeq \hat{\phi}$ and thus $r \left( \lambda = \infty \right) \simeq 1$. 
Therefore, the model exhibits a phase transition from a polarized to a consensus phase as $\lambda$ increases.

Eq.~\eqref{eq:fully_polarized} formalizes the intuition that multidimensional polarization is defined by individuals equally split into groups with opposite orientations $\hat{n}$. 
In addition, however, polarized initial opinions may be correlated. 
The angles formed by the orientations $\hat{n}$ determine the degree of correlation among initial opinions: intuitively, correlated opinions are orientated along similar directions and thus form small angles. 
We say then that topics are interdependent. 
Consequently, opinions are maximally correlated when the orientations $\hat{n}$ align along a single direction. 
In this situation, there are only two combinations of initial opinions, for instance, the unnormalized combination $(+1,+1, \ldots, +1)$ and the opposite $(-1,-1, \ldots, -1)$. 
That is, there exist only two groups of individuals, those with opinions equal to $+1$ for all topics, $x^{\left( \tau \right)} = +1$ $\forall \tau$, and those with opinions equal to $-1$ for all topics, $x^{\left( \tau \right)} = -1$ $\forall \tau$. 
These two groups must be composed of the same number of individuals, for Eq.~\eqref{eq:fully_polarized} to hold. 
The two combinations of opinions define thus two opposite ideologies, e.g., left vs right. 
When initial opinions are partially correlated, other combinations are allowed and orientations $\hat{n}$ are distributed in more than one dimension. 
In the case of uncorrelated opinions, instead,  
orientations $\hat{n}$ are equally distributed in all $D$ dimensions. We say then that topics are independent.

For the sake of simplicity, here we consider the simplest case in which initial opinions are represented by orthogonal orientations $\hat{n}$. 
In this way, we can avoid dealing with all possible angles formed by the orientations, a case we leave for future research. 
For the case of uncorrelated opinions, for $D$ topics we consider only $D$ possible orientations $\hat{n}$ and their opposite vectors $-\hat{n}$. 
Since the system is invariant \delete{for} \add{under} rotation with respect to any axis, without lack of generality we consider the orthonormal basis $\hat{e}_\tau$, $\tau=1, \ldots, D$, and its opposite $-\hat{e}_\tau$, as orientations for the uncorrelated case. 
Likewise, opinions are partially correlated if orientations span only some of the $D$ dimensions, while others do not contain any orientation $\hat{n}$. 
For instance, partially correlated opinions for $D = 4$ topics may be represented by orientations $\hat{n}$ lying in a plane, with two ``empty" dimensions. 
Fully correlated opinions are represented by a single vector and its opposite, as stated earlier. 
It is crucial to identify the correlations between initial opinions, since in principle the behavior of the order parameter varies depending on whether opinions are or not correlated.

\subsection{Mean-field solution}

At the mean-field level, where each individual interacts with every other, a general solution of the social compass model in $D$ dimensions can be found by the self-consistent equation $r = I(r, \hat{\phi})$ (see Appendix Section A, A-A), with
\begin{equation}
I(r, \hat{\phi}) =  \int \frac{P \left( \rho \right) P \left( \hat{n} \right) \left[ \lambda r + \rho \, ( \hat{n} \cdot \hat{\phi} ) \right]}{\sqrt{\left( \lambda r \right)^2 + 2 \lambda r \rho \, ( \hat{n} \cdot \hat{\phi} ) + \rho^2}} \, d\rho \, d\hat{n},
\label{eq:D-solution}
\end{equation}
where $\hat{\phi} = \vec{\phi}/r$ is the unit vector order parameter. 
From the previous equation, we can see that the model is completely defined by the probability distributions $P\left( \rho \right)$ and $P\left( \hat{n} \right)$, which we assume to be statistically independent. 
On the one hand, integrating the conviction $\rho$ can be quite complex. 
On the other hand, however, the dependence on the preferred orientation $\hat{n}$ is limited to the product $\hat{n} \cdot \hat{\phi}$.

The phase transition from a polarized to a consensus state is exhibited at a threshold value of the social influence $\lambda$. 
This threshold can be computed from Eq.~\eqref{eq:D-solution} as the instability point of the $r = 0$ solution, taking the form (see A-A1)
\begin{equation}
\lambda_c^\mathrm{MF} = \frac{1}{\int \frac{P \left( \rho \right)}{\rho} P \left( \hat{n} \right) | \hat{n} \times \hat{\phi} |^2 \, d\rho \, d\hat{n}}.
\label{eq:D-threshold}
\end{equation}

\begin{figure}[tbp]
     \centering
\includegraphics[width=0.9\columnwidth]{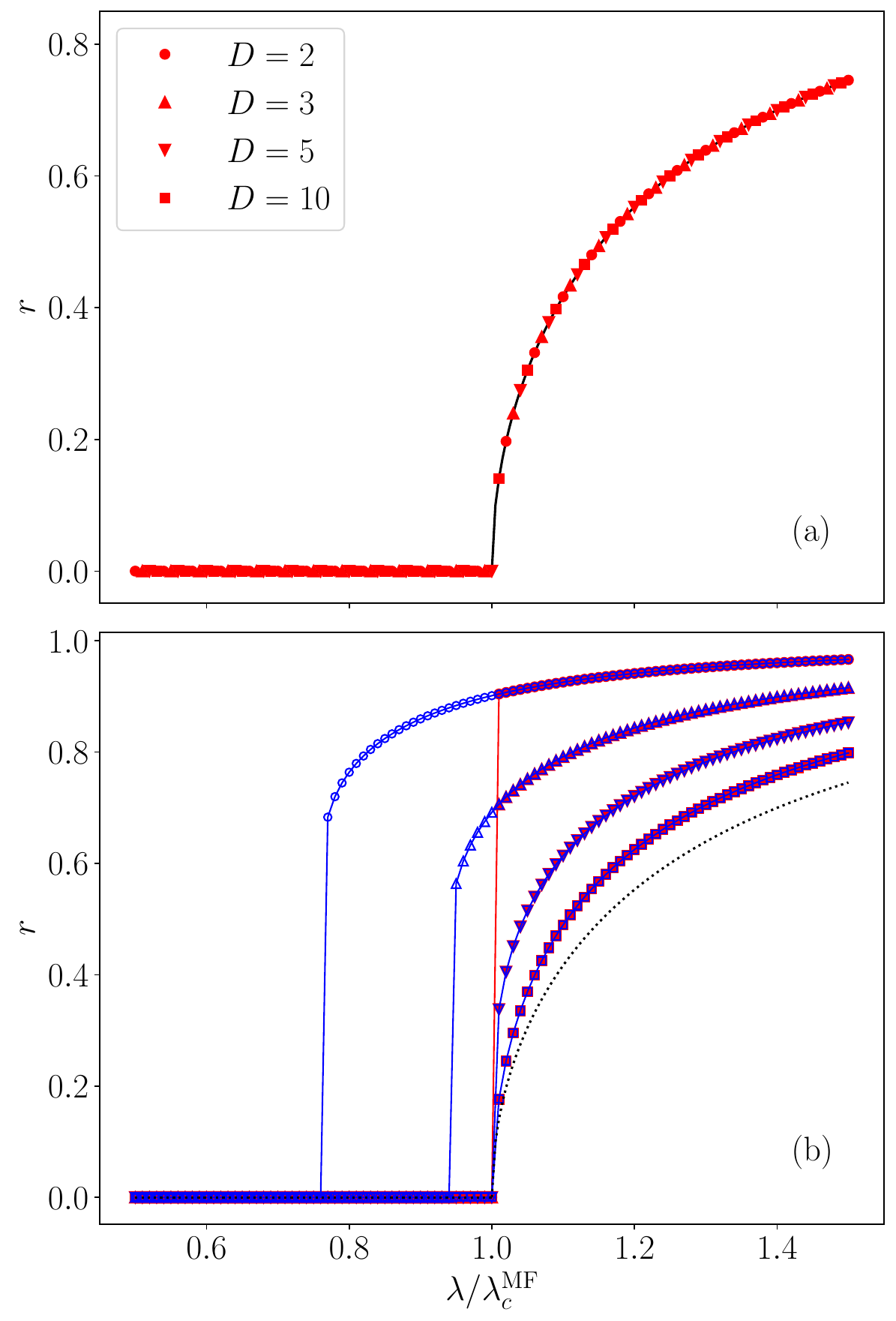}
    \caption{
    \textbf{Multidimensional depolarization dynamics at the mean-field level.} 
    Scalar order parameter $r$ as a function of the rescaled social influence $\lambda/\lambda_c^\mathrm{MF}$ for correlated (a) and uncorrelated (b) initial opinions. 
    Different dimensions $D$ are considered, and system size $N = 3 \times 10^6$. Conviction is fixed constant, $\rho = 1$. 
    Points (solid lines) represent numerical simulations (theoretical predictions). 
    For uncorrelated opinions, backward continuation (blue) is plotted in empty symbols. Curves of forward continuation (red) for different $D$ collapse for $\lambda \le \lambda_c^\mathrm{MF}$, while they are identical to backward continuations for $\lambda > \lambda_c^\mathrm{MF}$. 
    Dashed line (black) represents the solution for $D \to \infty$, which coincides with the one for correlated opinions.
    }
    \label{fig:multiD}
\end{figure}

In order to study analytically the behavior of the order parameter, here we consider the simplest case $P\left( \rho \right) = \delta \left( \rho - 1 \right)$, that is $\rho_i = 1$ $\forall i$. 
Indeed, this choice of a constant conviction of agents allows us to deduce the value of the dot product between $\hat{n}_i$ and $\hat{\phi}$ only from the correlations between initial opinions. 
Therefore, substituting it into Eq.~\eqref{eq:D-solution}, we obtain the self-consistent equation $r = I \left( r, D \right)$. 
Moreover, the modulus of the cross product $| \hat{n} \times \hat{\phi} |$ present in Eq.~\eqref{eq:D-threshold} can be determined from the dot product $\hat{n} \cdot \hat{\phi}$. 
In Ref.~\cite{ojer_modeling_2023}, where the social compass model was defined in $D = 2$, it was shown that, in the depolarized phase with $r > 0$, no initial opinion dominates over the remaining ones. 
This means that the absolute value of $\hat{n}_i \cdot \hat{\phi}$ has to be the same for each individual $i$. 
In the case of correlated opinions, in which some dimensions are empty, this can be achieved if and only if the average orientation $\hat{\phi}$ points in a direction perpendicular to all the preferred orientations, i.e., $\hat{n} \cdot \hat{\phi} = 0$. 
This then yields $| \hat{n} \times \hat{\phi} | = 1$. 
Thus, we find, for any dimension $D$, $I \left( r, D \right) = \lambda r / \sqrt{(\lambda r)^2 + 1}$, leading to the analytical solution $r \left( \lambda \right) = \sqrt{\lambda^2 - 1}/\lambda$, with $\lambda_c^\mathrm{MF} = 1$. 
Figure~\ref{fig:multiD}(a) confirms this, where the numerical order parameter obtained for different dimensions follows the same theoretical prediction, depicted as a black line.

In the case of uncorrelated opinions, however, preferred orientations are equally distributed in the $D$ dimensions, so there is no perpendicular direction to all of them. Therefore, no initial opinion prevails over the others if and only if $\hat{\phi}$ points in an arbitrary direction along one of the system diagonals, i.e., $\hat{n} \cdot \hat{\phi} = \pm 1/\sqrt{D}$. 
This leads to $| \hat{n} \times \hat{\phi} | = \sqrt{\left( D - 1 \right)/D}$. 
Unfortunately, an explicit solution from $r = I \left( r, D \right)$ cannot be provided. 
However, we can perform a Taylor expansion at $r = 0$ to predict the behavior of the order parameter in the vicinity of the phase transition. 
We find an upper critical dimension $D_c = 5$ for which the depolarization transition goes from being of first-order to second-order (see A-A2). 
Furthermore, for $D \ge D_c$, the resulting second-order phase transition shows the typical critical behavior $r\left( \lambda \right) \sim \left( \lambda - \lambda_c^\mathrm{MF} \right)^\beta$, with a threshold and an exponent $\beta$ depending on $D$ as
\begin{equation}
    \lambda_c^\mathrm{MF} \left( D \right) = \frac{D}{D - 1}, \quad \beta \left( D \right) = 
    \begin{cases}
        1/4 & \text{if } D = D_c \\
        1/2 & \text{if } D > D_c
    \end{cases}.
    \label{eq:critical_behavior}
\end{equation}
In Figure~\ref{fig:multiD}(b) we plot the numerical order parameter $r$ as a function of $\lambda/\lambda_c^\mathrm{MF}$ for different dimensions. 
We show in red (blue) the forward (backward) continuations, and their theoretical predictions obtained numerically in solid lines. 
In full agreement with our theoretical analysis, we observe that the depolarization transition is explosive with hysteresis if $D < D_c$, while continuous if $D \ge D_c$. 
Interestingly, in the limit $D \to \infty$ the transition exhibited by correlated opinions is recovered. Indeed, since $\hat{n} \cdot \hat{\phi} \to 0$ as $D$ increases, the average orientation becomes more and more orthogonal to all the preferred orientations. 
In Figure~\ref{fig:multiD}(b) we see that, for a large number of dimensions $D = 10$, the depolarization transition is very close to the solution for correlated opinions, represented by a dashed line.

\subsection{The two-dimensional social compass model}

In the previous Section, we showed the behavior of the model with respect to the number of dimensions $D$. 
We now focus on the simplest case of $D = 2$, i.e., two-dimensional opinion vectors towards two distinct topics.
In this case, the dynamics of the model can be expressed in the polar plane. 
The orientations are given by
\begin{equation}
  \hat{x}_i = \left( \cos \theta_i, \sin \theta_i \right),
  \label{eq:orientation}
\end{equation}
while their initial values are
\begin{equation}
  \hat{n}_i = \left( \cos \varphi_i, \sin \varphi_i \right).
  \label{eq:preferred}
\end{equation}
Considering the identity
\begin{eqnarray}
  \frac{d \theta_i}{d t} &=& \left( \cos \theta_i \right) ^2  \frac{d \theta_i}{d t} +  \left( \sin \theta_i \right) ^2  \frac{d \theta_i}{d t} \nonumber \\
  &=& \cos \theta_i \frac{d x_i^{(2)}}{d t} - \sin \theta_i \frac{d x_i^{(1)}}{d t}, \label{eq:identity}
\end{eqnarray}
we can substitute the derivatives of the components $x_i^{(1)}$ and $x_i^{(2)}$ from the model equation Eq.~\eqref{eq:model_orientation} into Eq.~\eqref{eq:identity}, leading to
\begin{equation}
    \dot{\theta}_i(t) = \lambda \sum_{j = 1}^N a_{ij} \sin \left[ \theta_j(t) - \theta_i(t) \right]
    + \rho_i \sin \left[ \varphi_i - \theta_i(t) \right],
    \label{eq:2dim-model}
\end{equation}
that is the two-dimensional social compass model~\cite{ojer_modeling_2023}.

It is interesting to compare the two-dimensional definition given by Eq.~\eqref{eq:2dim-model}, with the multidimensional social compass formulation given by Eq.~\eqref{eq:model_D}. 
The general multidimensional model is defined in terms of simple differences from opinion vectors. 
By imposing a constant modulus of the opinion vector in Eq.~\eqref{eq:model_D}, one obtains that the tendency of individuals to maintain their initial opinion is proportional to their conviction (i.e., agents with higher conviction are more stubborn), see Eq.~\eqref{eq:model_orientation}. 
Furthermore, the sine functions for both interaction terms in the two-dimensional case directly derive from the simple identity in Eq.~\eqref{eq:identity}. 
Therefore, the two-dimensional social compass model defined in Eq.~\eqref{eq:2dim-model} is fully equivalent to the general multidimensional case presented in Eq.~\eqref{eq:model_D}.


As we have shown above, the nature of the phase transition from polarization to consensus, as well as the critical exponent $\beta$, is always the same in the mean-field case, except for uncorrelated opinions for $D \le D_c = 5$. 
Therefore, in the following, we will focus on the simplest case of two-dimensional maximally correlated opinions. 
We thus consider a bimodal distribution of initial orientations,
\begin{equation}
P(\varphi) = \frac{1}{2} \left[ \delta(\varphi) + \delta(\varphi - \pi) \right],
\label{eq:angular}
\end{equation}
aimed at leading to a fully polarized initial state with $\vec{\phi}(0) = \vec{0}$. 
From Eq.~\eqref{eq:D-threshold} we find that, at the mean-field level, the phase transition from a polarized to a consensus state occurs at the threshold value
\begin{equation}
    \lambda_c^\mathrm{MF} = \left[ \int_0^\infty \frac{P(\rho)}{\rho} \, d\rho \right]^{-1}.
    \label{eq:threshold_MF}
\end{equation}
In the next Section, we will show analytically and numerically that, when a general social network mediates social influence among individuals in $D = 2$, the threshold value of the depolarization transition for correlated opinions can vanish, contrary to the mean-field case.

\section{Vanishing threshold on social networks}
\label{sec:vanish}

We consider the two-dimensional social compass model with maximally correlated initial opinions on a social network.
We compare two different approaches: the heterogeneous mean-field approximation and the perturbation theory.

\subsection{Heterogeneous mean-field approximation}

The simplest theory accounting for dynamical processes on networks is the heterogeneous mean-field (HMF) approximation~\cite{pv01a}. 
For uncorrelated networks~\cite{assortative}, the HMF approximation is equivalent to the so-called annealed network approximation~\cite{dorogovtsev07:_critic_phenom}, in which the adjacency matrix can be replaced by its annealed form $a_{ij} \simeq \frac{k_i k_j}{N \av{k}}$~\cite{Boguna09}. 
Within this approximation, one can characterize the depolarization transition in terms of a degree averaged global order parameter~\cite{ichinomiya_frequency_2004}.
In the two-dimensional case, in which opinion orientations of Eq.~\eqref{eq:orientation} can be expressed in the complex plane, a complex order parameter can be defined, taking the form
\begin{equation}
r(\lambda) \, \mathrm{e}^{i\psi} = \frac{1}{N \av{k}} \sum_{j=1}^N k_j \mathrm{e}^{i \theta_j(\lambda)},
\label{eq:global_HMF}
\end{equation}
where $\av{...}$ denotes the average value and $\psi$ measures the average orientation among the population. 
The threshold condition for the emergence of a depolarized phase can then be obtained by inserting Eq.~\eqref{eq:global_HMF} into Eq.~\eqref{eq:2dim-model} and considering the instability of the solution $r = 0$ in the self-consistent equation ensuing in the steady state. This leads to the HMF threshold (see A-B) 
\begin{equation}
\lambda_c^\mathrm{HMF} = \frac{\av{k}}{\av{k^2}} \lambda_c^\mathrm{MF},
\label{eq:threshold_HMF}
\end{equation}
where $\lambda_c^\mathrm{MF}$ corresponds to the mean-field threshold Eq.~\eqref{eq:threshold_MF}. 
This result, recovered also in epidemic spreading~\cite{boguna_epidemic_2002, boguna_absence_2003} and synchronization phenomena~\cite{ichinomiya_frequency_2004, lee_synchronization_2005}, indicates that, if the degree fluctuations are very large, the threshold can become negligibly small. 
In the particular case of scale-free networks~\cite{Barabasi:1999}, with a power-law degree distribution $P(k) \sim k^{-\gamma}$, for $\gamma \le 3$ the second moment $\av{k^2}$ diverges for an infinitely large network and therefore $\lambda_c^\mathrm{HMF}$ tends to zero. 
In the present modeling framework, this result implies that, if the underlying social network is characterized by such topology, any non-zero value of social influence can lead the system from polarization to consensus. 
On the other hand, for $\gamma > 3$ we obtain that $\av{k^2}$ converges to a constant and therefore $\lambda_c^\mathrm{HMF} > 0$. 
Thus, a finite threshold is obtained for weakly heterogeneous networks, even in the thermodynamic limit $N \to \infty$.

It is worth stressing that the depolarization threshold can vanish only if interactions are mediated by an (infinitely large) broad-tailed social network, while the threshold is finite for other network topologies and in particular in the mean-field case, when all individuals are connected all-to-all. 
The only exception to this general behavior holds for a uniform conviction distribution, $P(\rho) = 1$, defined in the interval $\left[ 0,1 \right]$, where according to Eq.~\eqref{eq:threshold_MF}, $\lambda_c^\mathrm{MF} = 0$. 
Indeed, at the mean-field level, the self-consistent equation Eq.~\eqref{eq:D-solution} reads
\begin{equation}
    r = \int_0^1 d\rho \, \frac{\lambda r}{\sqrt{(\lambda r)^2 + \rho^2}} = \lambda r \arctanh{ \left( \frac{1}{\sqrt{(\lambda r)^2 + 1}} \right)}. \label{eq:self_general}
\end{equation}
Apart from the solution $r = 0$, corresponding to the polarized state, we find the non-zero solution
\begin{equation}
    r(\lambda) = \frac{1}{\lambda \sinh{\left( \frac{1}{\lambda} \right)}},
    \label{eq:solution_uniform}
\end{equation}
which vanishes only for $\lambda = 0$, indicating a null threshold. The asymptotic behaviors of the order parameter are then
\begin{equation}
r \left( \lambda \right) \sim
    \begin{cases}
    	\frac{2}{\lambda}\mathrm{e}^{-1/\lambda} & \lambda \to 0 \\
        \frac{1}{1 + \frac{1}{6\lambda^2}} & \lambda \to \infty
    \end{cases}.
\label{eq:asymptotic_behavior}
\end{equation}
Interestingly, for any degree distribution $P(k)$, the case of a uniform conviction distribution leads to a self-consistent equation that follows as (see Eq.~\eqref{eq:hmf_selfconsistent})
\begin{equation}
    r = \frac{1}{\av{k}} \int k^2 P(k) \lambda r \arctanh{ \left( \frac{1}{\sqrt{(\lambda k r)^2 + 1}} \right)} \, dk \equiv f(r).
    \label{eq:self_alpha0}
\end{equation}
While the integral cannot be performed for a general $P(k)$, we can obtain the threshold as the value of $\lambda$ above which a non-zero solution exists. Using standard geometric arguments, we can see that a non-zero solution exists when the right-hand side of Eq.~\eqref{eq:self_alpha0} fulfills the condition $\left.f'(r) \right|_{r=0} \geq 1$. We have
\begin{equation}
    f'(r) = \frac{f(r)}{r} - \frac{1}{\av{k}} \int \frac{\lambda k^2 P(k)}{(\lambda k r)^2 + 1} \, dk.
\end{equation}
In the limit $r \to 0$ the term $f(r)/r$ diverges. 
Indeed, the argument of the $\mathrm{arctanh}$ in Eq.~\eqref{eq:self_alpha0} tends to $1$ and thus this function goes to infinity. 
Therefore, $\lim_{r \to 0} f'(r) = \infty$ for any network structure, implying that $\left.f'(r) \right|_{r=0}$ is larger than $1$ for any value of $\lambda$, which indicates that $\lambda_c^\mathrm{HMF} = 0$. 
A uniform conviction is then a sufficient condition to obtain a vanishing threshold for any $P(k)$, as it is also observed in the mean-field case.





\subsection{Perturbation theory}

Despite its appealing simplicity, the HMF theory cannot reproduce the behavior of the social compass model, as we will show below. 
Therefore, we move beyond the HMF approximation by using perturbation theory~\cite{restrepo_onset_2005}. 
One can tackle the two-dimensional model equation Eq.~\eqref{eq:2dim-model} \add{with a microscopic approach,} relying on the spectral properties of the adjacency matrix of the social network.
\add{This} method \delete{that} has been proved to provide a better description of some dynamical processes on networks, such as the susceptible-infected-susceptible model of epidemic spreading \add{within the so-called quenched mean-field theory}~\cite{Castellano2010, ferreira_epidemic_2012}. 
An analytical treatment of the model starts from the definition of a complex local order parameter~\cite{restrepo_onset_2005}
\begin{equation}
    r_j(\lambda) \, \mathrm{e}^{i\psi_j} = \sum_{\ell = 1} ^N a_{j\ell} \, \mathrm{e}^{i \theta_\ell (\lambda)},
    \label{eq:order_local}
\end{equation}
with $r_j$ being a local measure of order and $\psi_j$ the local average orientation, computed over the nearest neighbors of an agent $j$. 
The depolarized phase is then characterized by the local consensus among the agents, i.e., $r_j > 0$. 
Solving for the steady state, we can determine the perturbation threshold by letting $r_j \to 0$, which leads to (see A-C)
\begin{equation}
\lambda_c^\mathrm{P} = \frac{\lambda_c^\mathrm{MF}}{\Lambda_m},
\label{eq:threshold_QMF}
\end{equation}
where $\Lambda_m$ is the largest eigenvalue of the adjacency matrix. In the case of uncorrelated scale-free networks with a maximum degree $k_c$, this largest eigenvalue can be estimated as~\cite{chung_spectra_2003, Castellano2010}
\begin{equation}
\Lambda_m \simeq
    \begin{cases}
    	\av{k^2}/\av{k} & 2 < \gamma < 5/2 \\
        \sqrt{k_c} & \gamma > 5/2
    \end{cases}.
\label{eq:eigenvalue}
\end{equation}
Since $k_c$ grows as a function of $N$ for any $\gamma$~\cite{Boguna09, mariancutofss}, the consequence of Eq.~\eqref{eq:eigenvalue} is remarkable: in any uncorrelated random network with power-law distributed connectivities, the threshold value goes to zero as the network size goes to infinity. 
This is in contrast with the HMF approximation given by Eq.~\eqref{eq:threshold_HMF}, 
where a constant threshold was observed for $\gamma > 3$. 
\add{Within the perturbation theory, we can also obtain information about the local order parameter Eq.~\eqref{eq:order_local} in the vicinity of the critical point, which according to our results is proportional to the components of the principal eigenvector of the adjacency matrix (see Eq.~\eqref{eq:real_expansion}).}

In order to check the previous theoretical analysis, in the following we test the different predictions of both HMF approximation and perturbation theory, $\lambda_c^\mathrm{HMF}$ and $\lambda_c^\mathrm{P}$ respectively, on synthetic and empirical networks; see A-D for simulation details. 
For simplicity, in the rest of the paper we set $P(\rho) = \delta \left( \rho - 1 \right)$, i.e., $\rho_i = 1$ $\forall i$, which leads to $\lambda_c^\mathrm{MF} = 1$.


\begin{figure}[tbp]
     \centering
\includegraphics[width=0.9\columnwidth]{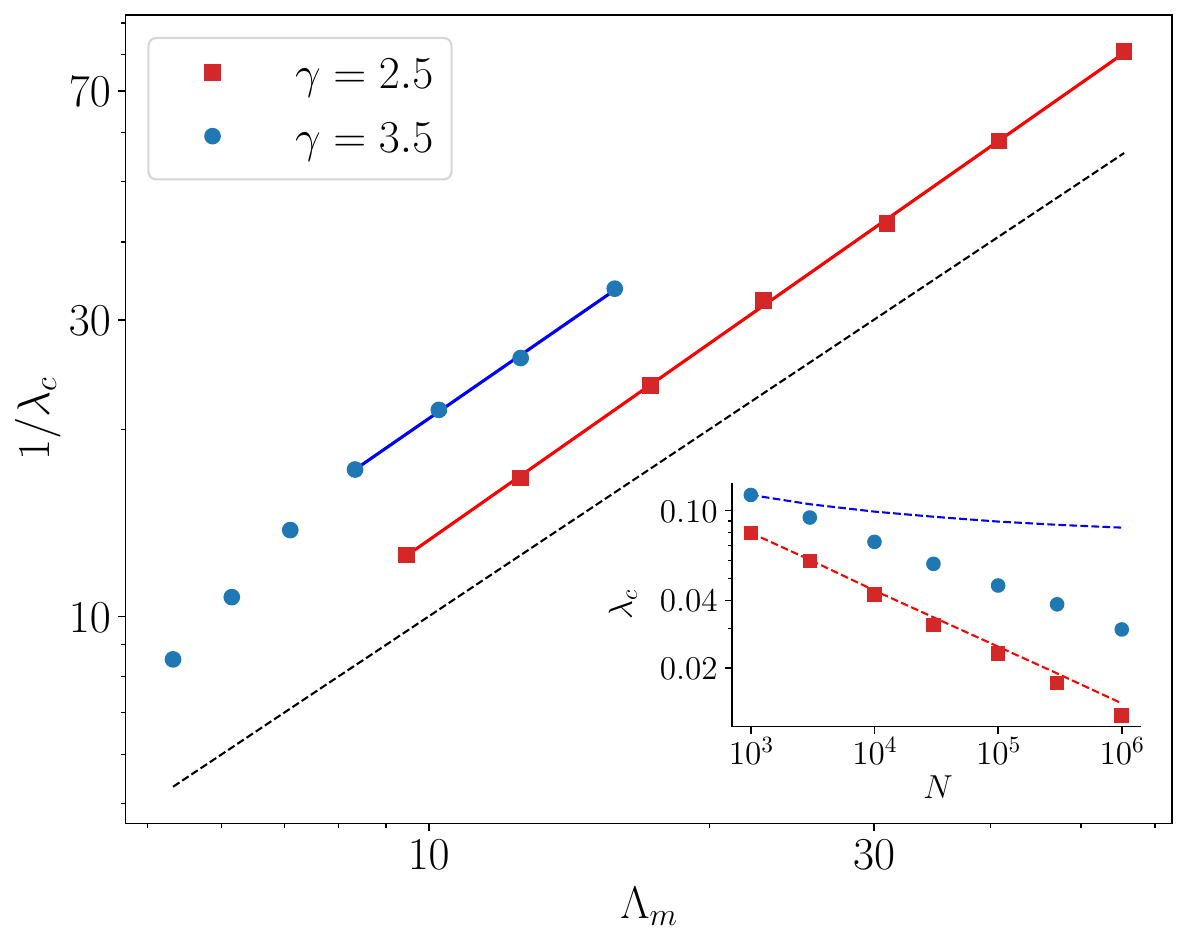}
    \caption{\textbf{Depolarization threshold on synthetic networks.} 
    Inset: Numerical threshold $\lambda_c$ as a function of size $N$ for scale-free synthetic networks with different exponents $\gamma$. The dashed lines, representing the theoretical prediction of the HMF theory, Eq.~\eqref{eq:threshold_HMF}, have been arbitrarily shifted to ease comparison with numerical data. 
    Main: Inverse of the numerical threshold $1/\lambda_c$ as a function of the largest eigenvalue $\Lambda_m$ for different values of $\gamma$. 
    The dashed line represents the theoretical prediction of the perturbation theory, Eq.~\eqref{eq:threshold_QMF}. Solid lines represent power regression $1/\lambda_c \sim \Lambda_m^a$, with $a = 1.05 \pm 0.01$ for $\gamma = 2.5$ and $a = 1.04 \pm 0.02$ for $\gamma = 3.5$. Due to finite-size effects, the power regression for $\gamma = 3.5$ is performed over the four largest network sizes. 
    Initial opinions: constant conviction $\rho = 1$, and bimodal orientation given by Eq.~\eqref{eq:angular}. 
    Each point is averaged over 100 UCM network realizations, being the eigenvalue of the adjacency matrix computed as the mean value of all the realizations. Error bars are smaller than symbols.
    }
    \label{fig:critical_point}
\end{figure}

\subsection{Synthetic networks}
\label{sec:vanish_UCM}

We build uncorrelated scale-free synthetic networks by using the uncorrelated configuration model (UCM)~\cite{Catanzaro05} with a minimum degree $k_\mathrm{min} = 3$ and maximum degree $k_c = \mathrm{min} \left( N^{1/2}, N^{1/(\gamma - 1)} \right)$~\cite{Boguna09,mariancutofss}. We consider degree exponents $\gamma = 2.5$ and $\gamma = 3.5$ to distinguish the two regimes (vanishing and non-vanishing threshold) of the HMF approximation. 
In Figure~\ref{fig:critical_point} (inset) we test the finite-size scaling of the depolarization threshold estimated by numerical simulations of the social compass model on scale-free UCM networks, by plotting $\lambda_c$ as a function of network size $N$. 
As we can see, the threshold decreases continuously for increasing $N$, for all values of $\gamma$. 
The theoretical HMF prediction, Eq.~\eqref{eq:threshold_HMF}, depicted as a dashed line, works quite well for $\gamma = 2.5$.  However, it predicts a constant threshold for $\gamma > 3$ in the thermodynamic limit, which is not compatible with numerical results for $\gamma = 3.5$.

Instead, numerical simulations are compatible with the perturbation theory, suggesting a vanishing threshold for large $N$ and for any $\gamma$. Indeed, in Figure~\ref{fig:critical_point} (main) we show the inverse of the depolarization threshold $1/\lambda_c$ as a function of the largest eigenvalue $\Lambda_m$, with the theoretical prediction of the perturbation theory, Eq.~\eqref{eq:threshold_QMF}, depicted as a black dashed line. 
As we can see, the numerical results are recovered with good accuracy by the theoretical expression, which predicts the correct scaling of the numerical threshold up to a constant multiplicative factor. We notice, however, that the agreement between theory and simulations is better for $\gamma = 2.5$ than for $\gamma = 3.5$, in which the perturbation theory prediction is recovered for the largest values of $\Lambda_m$, corresponding to large network sizes. 
Thus, we can attribute the deviation of the case $\gamma = 3.5$ from the theory to finite-size effects.

\subsection{Empirical networks}

\begin{figure}[tbp]
     \centering
\includegraphics[width=0.9\columnwidth]{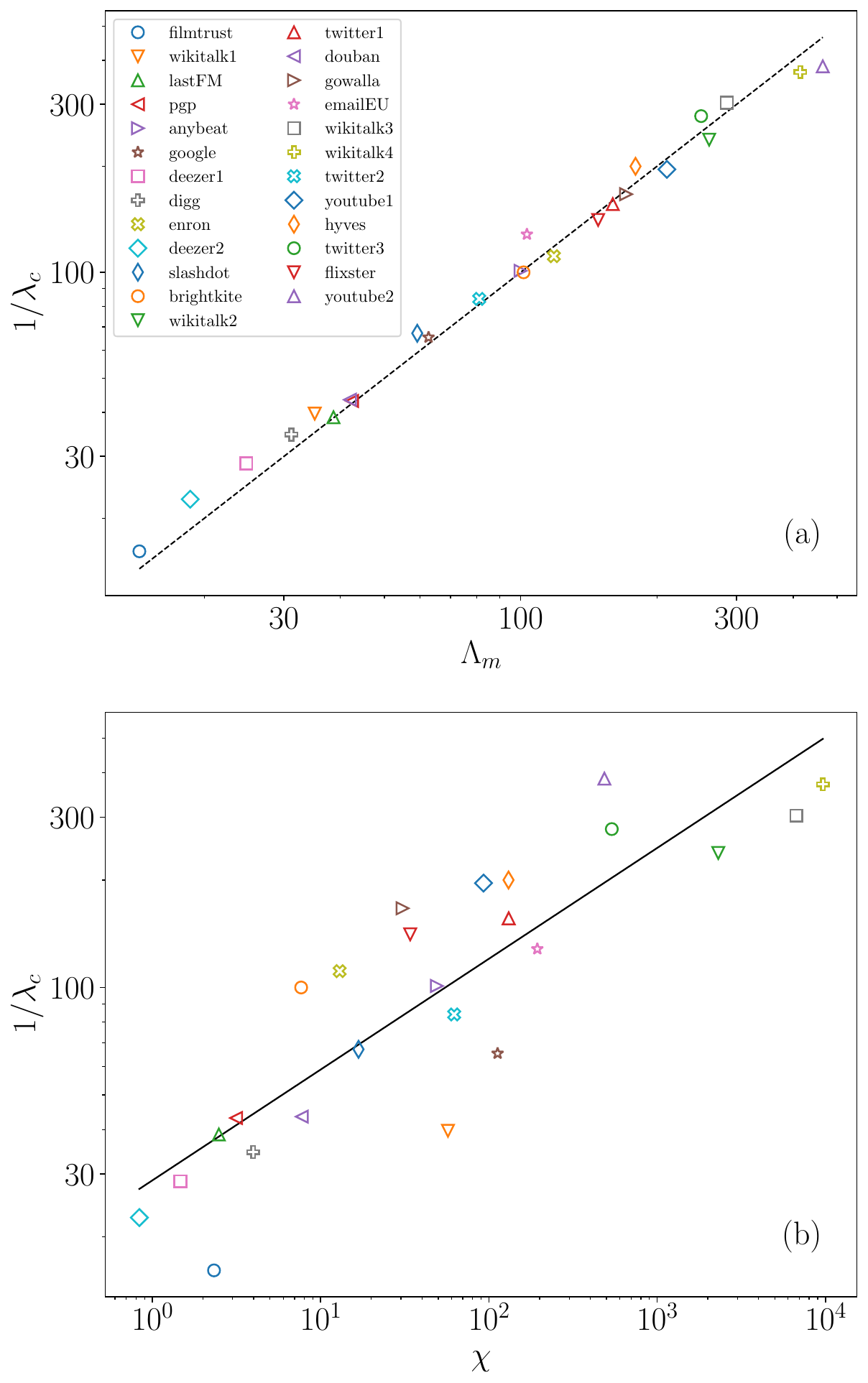}
    \caption{
    \textbf{Depolarization threshold on empirical social networks.} 
    (a) Inverse of the numerical threshold $1/\lambda_c$ as a function of the largest eigenvalue $\Lambda_m$ for different empirical networks. 
    The dashed line represents the theoretical prediction of the perturbation theory, Eq.~\eqref{eq:threshold_QMF}. 
    (b) Inverse of the numerical threshold $1/\lambda_c$ as a function of the heterogeneity parameter $\chi$ for different empirical networks. The solid line represents power regression $1/\lambda_c \sim \chi^a$, with $a = 0.31 \pm 0.03$ and a Pearson correlation coefficient $R = 0.87$. 
    Initial opinions: constant conviction $\rho = 1$, and bimodal orientation given by Eq.~\eqref{eq:angular}. 
    Each point is averaged over 10 initial orientation distributions. Error bars are smaller than symbols.
    }
    \label{fig:real_fits}
\end{figure}

To check the validity of our theoretical prediction in a more general scenario than the synthetic UCM model, we perform simulations of the social compass model on a set of 25 real social networks representing user-user interactions in online networking platforms, such as direct messages or friendship connections. Networks range in sizes between $N = 10^3$ and $N = 3 \times 10^6$ and have different levels of degree heterogeneity; see A-E for details. 
In Figure~\ref{fig:real_fits}(a) we present a scatter plot of the inverse of the numerically estimated thresholds $1/\lambda_c$ as a function of the largest eigenvalue $\Lambda_m$ computed for each network. 
As we can see, the perturbation theory provides an excellent prediction of the threshold value for all the different empirical networks considered, confirming the linear scaling, Eq.~\eqref{eq:threshold_QMF}, depicted as a black dashed line.

Noteworthy, the largest eigenvalue is closely related to degree heterogeneity~\cite{nishikawa_heterogeneity_2003, arenas_synchronization_2008}: the more heterogeneous the network, the larger the largest eigenvalue. 
Hence, one can characterize the onset of opinion depolarization according to the network heterogeneity. To do so, we compute the heterogeneity parameter $\chi = \av{k^2}/\av{k}^2 - 1$ in order to quantify how heterogeneous are the different empirical networks. 
Figure~\ref{fig:real_fits}(b) shows the inverse of the depolarization threshold $1/\lambda_c$ obtained numerically as a function of $\chi$. 
We see that 
the threshold value decreases as the degree heterogeneity increases, in agreement with what we obtained by using synthetic networks. A linear regression in logarithmic scale is shown as a black solid line, with a significant Pearson coefficient. 

\section{Network heterogeneity mitigates polarization}
\label{sec:hubs}

\begin{figure}[tbp]
     \centering
\includegraphics[width=0.9\columnwidth]{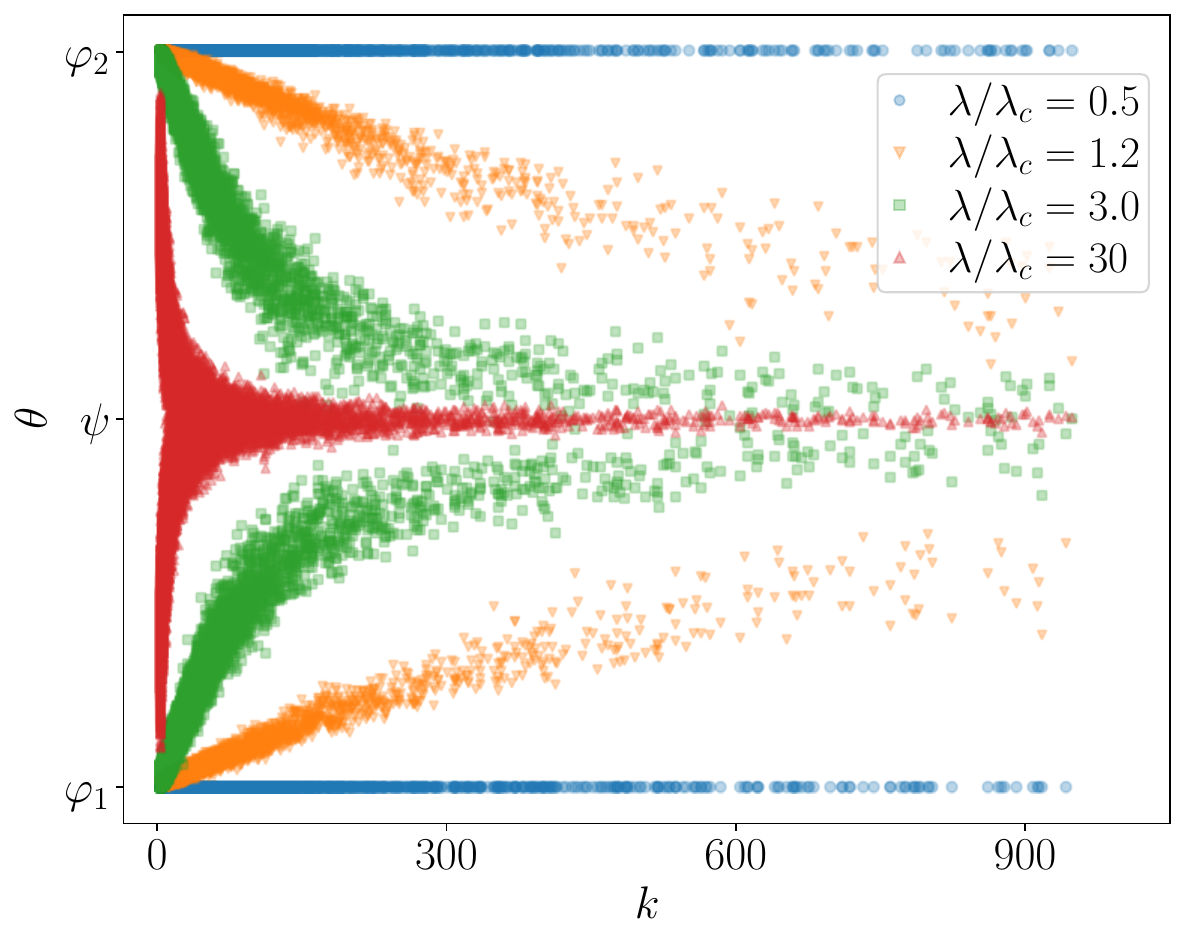}
    \caption{
    \textbf{Hubs promote consensus in polarized social networks.} 
    Stationary opinion orientation $\theta$ as a function of the degree $k$ of the nodes, for different values of the rescaled social influence $\lambda/\lambda_c$. 
    A highly heterogeneous scale-free UCM network with degree exponent $\gamma = 2.1$ and size $N = 10^5$ is used. 
    Initial opinions: constant conviction $\rho = 1$, and bimodal orientation given by Eq.~\eqref{eq:angular}, thus $\varphi_1 = 0$ and $\varphi_2 = \pi$ are the preferred orientations, while $\psi = \pi/2$ is the consensus.
    }
    \label{fig:consensus_degree}
\end{figure}

A smaller threshold for increasingly heterogeneous networks can be understood by the presence of highly connected nodes (hubs). 
\delete{Hubs, indeed, extend their social influence to a large number of nodes, while the social influence they receive from those neighbors is averaged down.} 
\add{Hubs have more connections and thus can explore the average opinion of the network, corresponding to the consensus, more efficiently. 
Once they have locked their opinion in this state, 
they can pull the opinion of other nodes to this value, due to their large number of connections. 
A similar effect is observed in flocking models on social networks~\cite{miguel_effects_2018}.} 
Therefore, the presence of hubs promotes consensus formation, decreasing the depolarization threshold. 
Figure~\ref{fig:consensus_degree} illustrates this observation by showing the final orientation $\theta$ as a function of the degree $k$ of nodes of a very heterogeneous scale-free UCM network with $\gamma = 2.1$. 
We indicate as $\varphi_1$ and $\varphi_2$ the two opposite preferred orientations of the nodes, while
$\psi$ indicates the consensus reached. 
When $\lambda < \lambda_c$ (blue points), all nodes align to their preferred orientation $\theta = \varphi_1, \varphi_2$. 
As soon as $\lambda > \lambda_c$ (orange points), a consensus starts to emerge, the larger the degree of a node the closer its orientation to consensus. 
As $\lambda$ increases (green points), the hubs almost reach consensus, while low-degree nodes remain at their preferred orientation $\theta = \varphi_1, \varphi_2$. 
For $\lambda \gg \lambda_c$ (red points), all highly connected nodes reach consensus $\theta = \psi$, while some nodes with small degrees still do not.

To fully test the effects of hubs on the depolarization process, we develop an algorithm that increases the heterogeneity of the underlying network, while opinion dynamics unfolds on it. 
Recent studies have investigated how the formation of social ties affects opinion polarization. 
Many of them are based on opinion similarity, in which agents tend to connect with those that hold similar beliefs~\cite{baumann19, ferraz_de_arruda_modelling_2022, liu_emergence_2023}. 
Others, instead, consider only the structural properties of the network, like link recommendation algorithms~\cite{santos_link_2021} and behavioral influence processes~\cite{centola_spread_2010}. 
Here we focus on this latter group, relying solely on the structure of the social network to characterize depolarization. 
We explore two alternatives, by using either global information or only local information about the network connectivity. 

In both cases, we start from a random regular network with $k_i = 10$ $\forall i$, $N = 10^4$, and rewire the edges between nodes increasing the heterogeneity of the network topology. 
We fix the minimum degree $k_\mathrm{min} = 3$ and define the maximum as $k_\mathrm{max} = \mathrm{max} \left( \{ k_1, \dots , k_N \} \right)$. 
Inspired by Ref.~\cite{whigham_network_2016}, a single step of the rewiring algorithm is defined by the following 3 stages, in which a low degree node is rewired to a high degree node, increasing thus the maximum degree of the network. 
The amount of information used about the network connectivity (global or local) differentiates the two algorithms in the second stage.

\begin{enumerate}
    \item \textbf{Giver node:} 
    Randomly select a giver node $i$, where $k_i > k_\mathrm{min}$. If $k_i = k_\mathrm{max}$, ensure that node $i$ is not the only one with maximum degree.
    \item \textbf{Receiver node:} 
    Two different procedures: 
    \begin{itemize}
        \item \textbf{Global information Algorithm:} Preferentially select a receiver node $j$ with probability
        \begin{equation}
        P \left( j, i \right) = \frac{k_j \left[ j \neq i, k_j \ge k_i \right]}{\sum_{\ell = 1}^N k_\ell \left[ \ell \neq i, k_\ell \ge k_i \right]}.
    \label{eq:preferential}
    \end{equation}
        \item \textbf{Local information Algorithm:} Randomly choose a node and select its largest degree neighbor $j$ as the receiver node. Ensure that $j \neq i$ and $k_j \ge k_i$.
    \end{itemize}
    \item \textbf{Rewiring:} 
    Select the smallest degree node $q$ connected to $i$. If $q \neq j$ and $q$ is not connected to $j$, rewire edge $e_{iq}$ to node $j$ by creating $e_{qj}$, otherwise return to stage 1.
\end{enumerate}

\begin{figure}[tbp]
     \centering
\includegraphics[width=0.9\columnwidth]{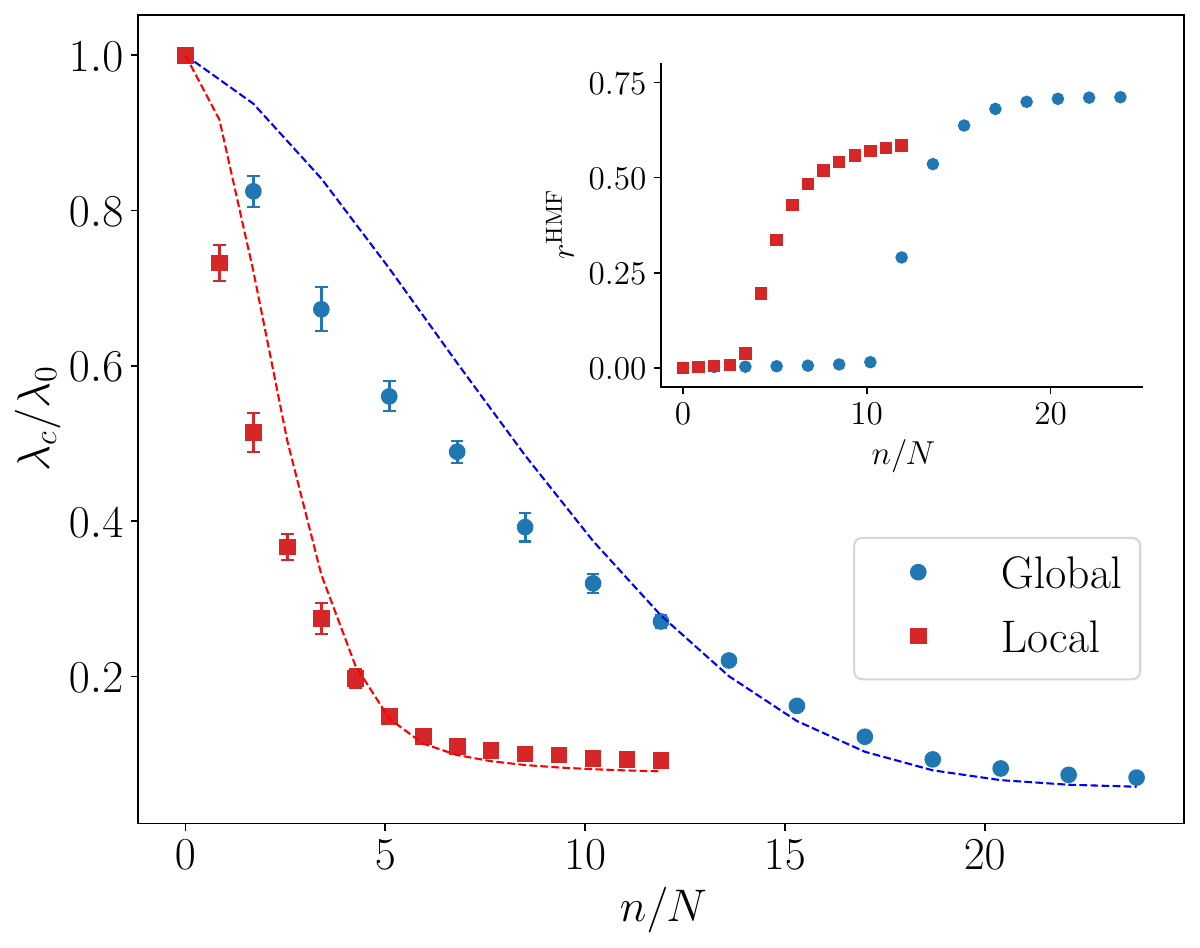}
    \caption{
    \textbf{Increasing network heterogeneity promotes depolarization.} 
    Main: Ratio of numerical thresholds $\lambda_c/\lambda_0$ as a function of the attempted rewiring step divided by the network size $n/N$. The dashed lines represent the theoretical prediction of the perturbation theory, Eq.~\eqref{eq:threshold_QMF}. 
    Inset: Numerical degree averaged order parameter $r^\mathrm{HMF}$ as a function of $n/N$ at a fixed social influence $\lambda = 0.03$. 
    The simulations are performed by using both global (blue) and local (red) information about the network connectivity, with a system size $N = 10^4$. 
    Initial opinions: constant conviction $\rho = 1$, and bimodal orientation given by Eq.~\eqref{eq:angular}. 
    Each point, with the corresponding error bar, is averaged over 10 different rewirings.
    }
    \label{fig:RtoS}
\end{figure}

Figure~\ref{fig:RtoS} (main) shows the ratio between the depolarization threshold obtained for the rewired network, $\lambda_c$, and the one obtained by using the initial homogeneous network, $\lambda_0$, as a function of the ratio between the number of attempted rewiring steps, $n$, and the network size, $N$. 
We see that as the edges are rewired, the heterogeneity increases and the threshold value diminishes until reaching a plateau. 
This observation confirms that the network heterogeneity enhances depolarization, being the presence of hubs key to fostering consensus. 
For highly heterogeneous connectivities, furthermore, the numerical threshold is predicted with high accuracy by the perturbation theory Eq.~\eqref{eq:threshold_QMF}, depicted as dashed lines. 
Interestingly, if the rewiring is performed by using only local information, the network becomes more heterogeneous at a faster rate than by using global information, requiring a lower number of rewiring steps to depolarize the system.

In addition, Figure~\ref{fig:RtoS} (inset) shows the degree averaged order parameter $r^\mathrm{HMF}$ defined in Eq.~\eqref{eq:global_HMF} as a function of $n/N$ for a fixed social influence $\lambda = 0.03$. 
We see that the system reaches consensus only if the network is sufficiently heterogeneous, otherwise it remains polarized. 
In particular, the local algorithm seems to be the fastest option to depolarize the system, instead of rewiring edges through a preferential selection following Eq.~\eqref{eq:preferential}.

\section{Opinion-degree correlations hinder depolarization}
\label{sec:patrizi}


Next, 
we study how the correlations between the initial opinions of agents and their degree affect the depolarization dynamics. 
To this aim, we run the model by assigning the orientations $\hat{n}_i$ to the nodes depending on their degree $k_i$. 
Specifically, we consider that highly connected nodes share the same initial opinions, while lowly connected individuals share the opposite ones. 
We quantify the amount of opinion-degree correlation in the network by the parameter $\mu$, indicating the fraction of nodes 
whose orientation is assigned depending on their degree. 
For instance, with $\mu = 0.5$, the 25\% of individuals with the largest degree have one preferred orientation, the 25\% with the lowest degree have assigned the opposite orientation, while the remaining 50\% of individuals hold a randomly chosen orientation between the two. 
We test the effect of these correlated initial conditions on both synthetic and empirical networks.

\begin{figure}[tbp]
     \centering
    \includegraphics[width=0.9\columnwidth]{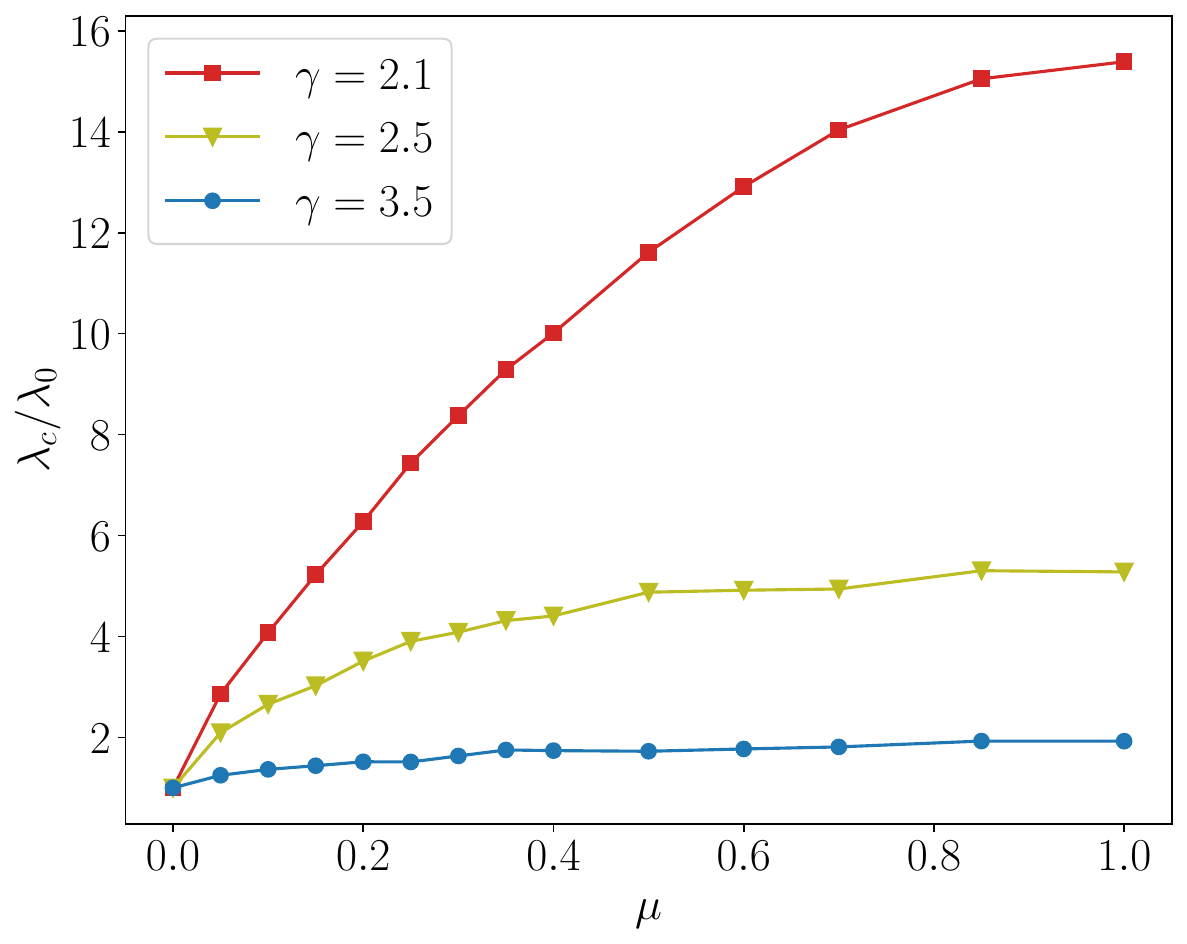}
    \caption{
    \textbf{Correlations between initial opinions and degree hinder depolarization in synthetic networks.} 
    Ratio of numerical thresholds $\lambda_c/\lambda_0$ as a function of the percentage $\mu$ for scale-free synthetic networks with different exponents $\gamma$. 
    Initial opinions: constant conviction $\rho = 1$, and bimodal orientation given by Eq.~\eqref{eq:angular}. 
    Each point is averaged over 25 UCM network realizations with $N = 10^5$. Error bars are smaller than symbols.
    }
    \label{fig:patrizi}
\end{figure}

\subsection{Synthetic networks}

Similarly to Section~\ref{sec:vanish_UCM}, we consider scale-free UCM networks with different exponents $\gamma$. 
Figure~\ref{fig:patrizi} shows the ratio between the depolarization threshold obtained for correlated initial conditions, $\lambda_c$, and the one obtained without opinion-degree correlations, $\lambda_0$, as a function of the fraction of individuals with correlations between initial orientation and degree, $\mu$. 
As we can see, the more individuals with initial opinion correlated with their degree (larger $\mu$), the larger the threshold value, up to roughly 15 times larger. 
As expected, this relationship strongly depends on the structural heterogeneity of the network, quantified by the exponent of the power-law degree distribution $\gamma$. 
The increasing tendency of the depolarization threshold is more pronounced for low values of $\gamma$ (strongly heterogeneous networks), while the increase is smaller for large $\gamma$ (more homogeneous networks) and quickly reaches a plateau for high $\mu$. 
\add{These observations can be explained in terms of the effect of hubs. When initial opinions are correlated to degrees, high-degree nodes have the same opinion and are also mostly connected to each other, in the rich club effect~\cite{colizza_detecting_2006}. Therefore, the neighborhood they gauge has essentially the same opinion, and it takes a much larger social influence to overcome polarization and render the system into consensus.} 
\delete{In the former case} \add{In the case of small $\gamma$}, the presence of very large hubs with the same preferred orientation dominates the dynamics and strongly hinders depolarization. \delete{In the latter case} \add{When $\gamma$ is large}, on the other hand, hubs are smaller and fewer, thus the effect on the depolarization transition is negligible.


\subsection{Empirical networks}

\begin{figure}[tbp]
     \centering
    \includegraphics[width=0.9\columnwidth]{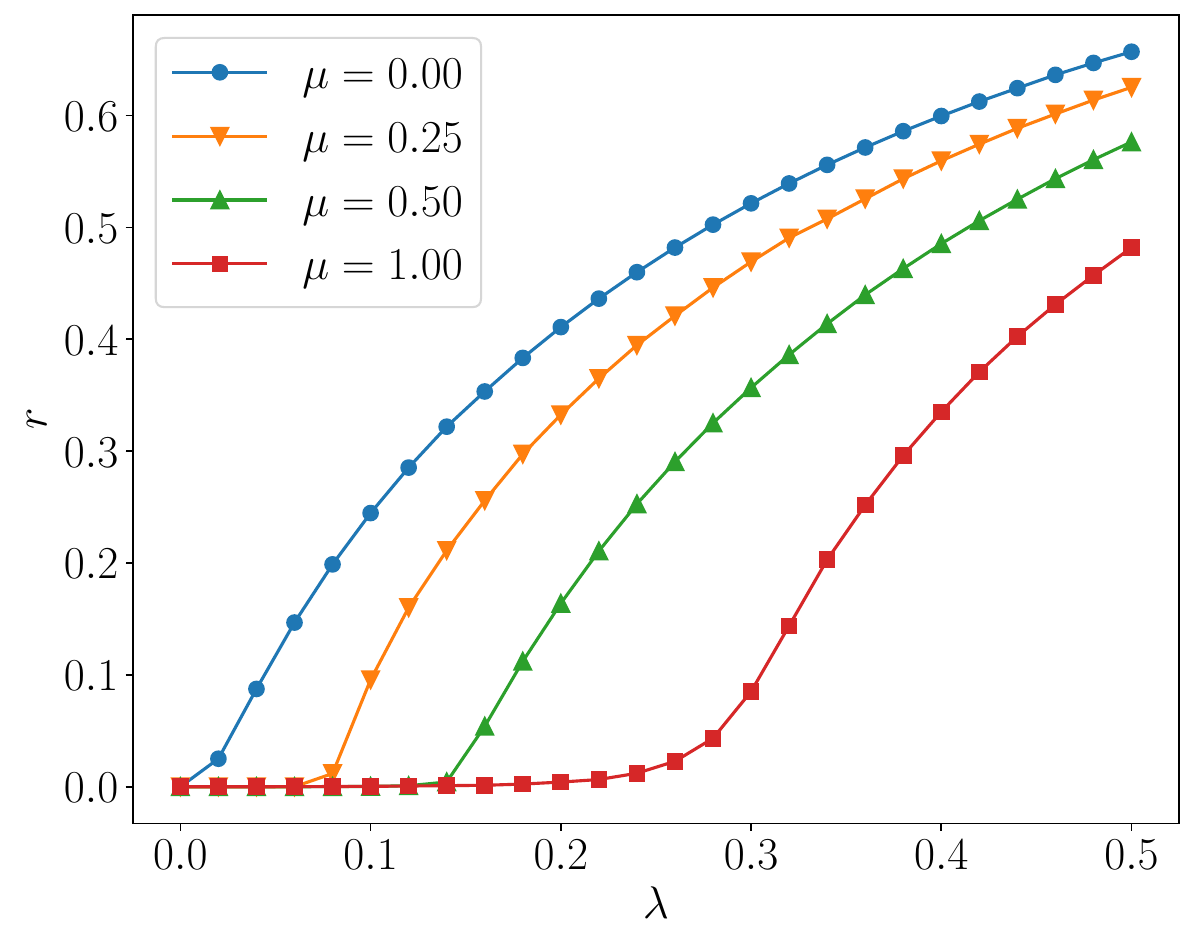}
    \caption{
    \textbf{Correlations between initial opinions and degree hinder depolarization in empirical networks.} 
    Scalar order parameter $r$ as a function of the social influence $\lambda$ for the empirical network \texttt{brightkite}, $N = 58228$, for different values of $\mu$. 
    Initial opinions: constant conviction $\rho = 1$, and bimodal orientation given by Eq.~\eqref{eq:angular}.
    }
    \label{fig:patrizi_real}
\end{figure}

The results found for synthetic networks hold also when the underlying network is a real social network. 
Figure~\ref{fig:patrizi_real} shows the scalar order parameter $r$ defined in Eq.~\eqref{eq:scalar_order} as a function of the social influence $\lambda$ for the real network \texttt{brightkite}. 
We see that, as $\mu$ increases, greater values of $\lambda$ are needed for the emergence of a consensus state with $r > 0$. 
Opinion-degree correlations are thus detrimental for depolarizing the system, while when opinions are fully independent of social connectivity ($\mu = 0$) consensus is reached for much smaller values of $\lambda$. 
In Supplemental Material we show equivalent results for a different empirical network.

Therefore, experiments with real and synthetic networks reveal that when initial (preferred) opinions are correlated with the degree, depolarization is much harder to achieve. 
In social networks, the degree, i.e., the number of different social connections of an individual, is related to popularity and social status, the larger the degree, the more popular and probably influential the individual is. 
If popular agents share the same opinion, opposite to the opinion of unpopular ones, reaching consensus is much more difficult and the network remains polarized for much higher values of social influence. 
Interestingly, correlations between opinions and social status of individuals have been reported in the last years by several studies, arguing that people from different social classes hold distinct positions on polarized issues like civil liberties, environmental protection or immigration~\cite{ares_issue_2022}. 
Moreover, this social classification is often strongly linked with the economic status, which substantially contributes to shaping political ideology~\cite{brown-iannuzzi_subjective_2015}.

\section{Conclusions}
\label{sec:discussion}


In this paper, we proposed the multidimensional social compass model, an opinion depolarization model in which opinions held by agents vary across multiple topics. 
The depolarization process is based on only two key ingredients: i) DeGroot learning, driven by the social influence exerted by individuals~\cite{degroot_reaching_1974}, and ii) preference of agents to maintain their initial opinions, as formalized by the Friedkin-Johnsen model~\cite{friedkin_social_1990}. 
Due to the competition between both interaction terms, the model exhibits a phase transition from a polarized to a consensus state at a threshold value of the social influence. 
The nature of such depolarization transition depends on the correlations between the initial opinions of agents. 
Within this modeling framework, we explored under which conditions the depolarization dynamics is promoted or hindered.

First, we studied the effect of dimensions (representing the number of topics discussed at the same time) in the simplest case of mean-field approximation and agents with constant conviction $\rho = 1$. 
We found that, if opinions are correlated, the transition is always continuous and the threshold is independent of the number of topics. 
Uncorrelated opinions, instead, trigger a first-order phase transition if the number of topics is lower than $D = 5$, otherwise the transition is of second-order. 
Moreover, the depolarization threshold decreases with the number of topics, and tends to the one obtained in the correlated case in the limit $D \to \infty$. 
Therefore, when the initial opinions of agents are uncorrelated and the number of topics discussed is low, the transition from depolarization to consensus is harder, i.e., it occurs for larger values of social influence.

Next, we investigated the effects of a nontrivial connectivity among the agents, i.e., the structure of the social network, for correlated opinions and the simplest case $D = 2$. 
We found that the depolarization threshold is fully determined by the topology of the underlying network, in agreement with perturbation theory. 
As a result, the threshold value can be very small, vanishing within the limit of a large network size. 
This indicates that even a weak social influence among individuals may be able to trigger consensus in a polarized, heterogeneous social network. 
This behavior is recovered in other networked dynamical processes, like epidemic spreading~\cite{boguna_absence_2003, Castellano2010}. 
In other cases, by contrast, a non-zero threshold is observed. For example, the synchronization process of coupled oscillators in scale-free networks shows a positive threshold for a degree exponent $\gamma > 3$, as predicted by HMF theory~\cite{peron_onset_2019}.

Since most empirical social networks are highly heterogeneous, our results imply that it could be easier to reduce opinion polarization in these realistic settings. 
This is due to the presence of hubs, which promote consensus by decreasing the strength of social influence needed to depolarize the system. 
Therefore, the more heterogeneous the underlying social network, the smaller the expected depolarization threshold. 
This last finding has interesting implications for real social media: the actions of very popular users (hubs) are pivotal to achieving depolarization, as they can reach consensus earlier than the rest of the network. 
To test this observation, we designed a rewiring algorithm that increases the structural heterogeneity of a synthetic network. 
By running the social compass model on top of it, we observed that indeed the depolarization threshold decreases as the network heterogeneity increases.

Finally, we showed 
that if initial opinions are correlated with the degree of nodes, the depolarization dynamics is hindered. 
As expected, this effect is much more pronounced if the network is very heterogeneous, with several very large hubs. 
In that case, the amount of social influence to depolarize the population increases very significantly. 
Since the degree of individuals is related to their popularity and social status, our results show that if popular agents share the same opinions, opposite to unpopular agents, depolarizing the network is much more difficult.

Our theoretical formulation, however, could embrace a more general framework. 
To keep the analytical treatment simple and the paper as readable as possible, in this work we explored limited scenarios, making some simplifying assumptions for different cases. 
First, we considered that preferred opinions are only represented by orthogonal orientations. 
As a consequence, correlations between opinions arise if orientations are not distributed in all $D$ dimensions. 
However, correlations would also appear if they are equally distributed in the different dimensions but forming small angles along a certain direction instead of being orthogonal. 
A more sophisticated mathematical treatment is required then, which should be consistent with our analysis shown here.

Additionally, we provided analytical results only for the case of constant conviction of agents. More complex and realistic conviction distributions could be considered to study the effects of stubbornness on multidimensional depolarization dynamics. 
Lastly, in light of how the dimensionality affects the depolarization transition in the mean-field case, we decided to restrict our analysis concerning the impact of the structure of the underlying network only to the two-dimensional case of fully correlated topics. 
Future work should be devoted to validating if these findings generalize to more dimensions. 

\add{Finally, we did not include in our model some realistic features that could alter its behavior.
For instance, external fields, i.e., the social influence exerted not by peers but media, both traditional (newspapers or TV) and online, could play an important role.
If such an external field is not homogeneous in the population, it could tilt the balance towards the polarized state~\cite{fletcher_how_2020}. 
Another important factor could be the presence of zealots: individuals that never (or hardly) change their opinion whatever social influence they receive. 
Zealotry, represented by present-day political pundits, coupled with unidirectional information transfer, could make it impossible to reach any sort of consensus, whatever the network structure, especially if zealots are very well connected. 
Future venues for research include studying the effects of social media and zealotry on depolarization dynamics.}

It is worth remarking that, despite these simplifying assumptions adopted to analytically treat the problem, 
our findings are valid for a remarkably general multidimensional depolarization model. 
The model is indeed formulated by choosing the simplest linear functional form for both social influence and stubbornness, i.e., differences between opinion vectors. 
Interestingly, by imposing a constant modulus of opinion vectors, the tendency of individuals to maintain their initial orientations becomes proportional to the conviction. 
Despite such a simple theoretical framework, the model shows a very rich phenomenology, clearly indicating several conditions that can promote the depolarization dynamics, such as the presence of correlations between the topics discussed, the structural heterogeneity of the underlying social network, or the absence of correlation between the popularity of individuals and their preferred opinions. 
Future research should be dedicated to collecting observational data from social networks and, more importantly, designing experiments to test some of these theoretical insights.

\begin{acknowledgments}
    We acknowledge financial support from the Spanish MCIN/AEI/10.13039/501100011033, under Projects No. PID2019-106290GB-C21 and No. PID2022-137505NB-C21.
\end{acknowledgments}

\appendix

\section{Mean-field solution of the multidimensional social compass model}

At the mean-field level, the model equation Eq.~\eqref{eq:model_orientation} can be written as
\begin{equation}
  \frac{d \hat{x}_i}{d t} = \frac{\lambda}{N} \sum_j \left[ \hat{x}_j - (\hat{x}_j \cdot \hat{x}_i) \, \hat{x}_i \right] + \rho_i \left[ \hat{n}_i - (\hat{n}_i \cdot \hat{x}_i) \, \hat{x}_i \right].
  \label{eq:model_MF}
\end{equation}
In terms of the vector order parameter, the previous equation reads
\begin{equation}
  \frac{d \hat{x}_i}{d t} = \lambda \left[ \vec{\phi} - (\vec{\phi} \cdot \hat{x}_i) \, \hat{x}_i \right] + \rho_i \left[ \hat{n}_i - (\hat{n}_i \cdot \hat{x}_i) \, \hat{x}_i \right].
  \label{eq:model_order}
\end{equation}
In the steady state, the time derivative equals to zero. Rearranging some terms, we find the condition
\begin{equation}
    \lambda \vec{\phi} + \rho_i \hat{n}_i = 
    \left[ \lambda \, \vec{\phi} \cdot \hat{x}_i + \rho_i \, \hat{n}_i \cdot \hat{x}_i \right] \hat{x}_i.
    \label{eq:steady_condition}
\end{equation}
Importantly, moreover, we observe that
\begin{equation}
    r^2 = \vec{\phi} \cdot \vec{\phi} = \vec{\phi} \cdot \frac{1}{N} \sum_i \hat{x}_i = \frac{1}{N} \sum_i \vec{\phi} \cdot \hat{x}_i,
    \label{eq:order_phi}
\end{equation}
so the mean-field solution of the model will be given by an expression of $\vec{\phi} \cdot \hat{x}_i$ in terms of the system variables $\lambda$, $\rho_i$ and $\hat{n}_i$. 
For convenience, let us define first the dot products
\begin{equation}
    \alpha_i \equiv \hat{\phi} \cdot \hat{x}_i, \quad 
    \beta_i \equiv \hat{\phi} \cdot \hat{n}_i, \quad 
    \gamma_i \equiv \hat{n}_i \cdot \hat{x}_i,
    \label{eq:dot_products}
\end{equation}
where $\hat{\phi} = \vec{\phi}/r$. Squaring both sides of Eq.~\eqref{eq:steady_condition}, we find
\begin{equation}
    \lambda^2 r^2 + 2 \lambda \rho_i r \beta_i + \rho_i^2 = \left( \lambda r \alpha_i + \rho_i \gamma_i \right)^2.
    \label{eq:steady_square}
\end{equation}
On the other hand, multiplying both sides of Eq.~\eqref{eq:steady_condition} by $\vec{\phi}$, we find
\begin{equation}
    \lambda r^2 + \rho_i r \beta_i = \left( \lambda r \alpha_i + \rho_i \gamma_i \right) r \alpha_i.
    \label{eq:steady_order}
\end{equation}
From the previous equation, we write
\begin{equation}
    \rho_i \gamma_i = \frac{\lambda r + \rho_i \beta_i}{\alpha_i} - \lambda r \alpha_i.
    \label{eq:gamma}
\end{equation}
Substituting it into Eq.~\eqref{eq:steady_square}, we find
\begin{equation}
    \left( \lambda r \alpha_i \right)^2 + 2 \lambda \rho_i r \beta_i \alpha_i^2 + \rho_i^2 \alpha_i^2 = \left( \lambda r + \rho_i \beta_i \right)^2,
\end{equation}
which leads to
\begin{equation}
    \alpha_i = \frac{\lambda r + \rho_i \beta_i}{\sqrt{\left( \lambda r \right)^2 + 2 \lambda \rho_i r \beta_i + \rho_i^2}}.
    \label{eq:alpha}
\end{equation}
From Eq.~\eqref{eq:order_phi}, we then write
\begin{equation}
    r = \frac{1}{N} \sum_i \alpha_i = \frac{1}{N} \sum_i \frac{\lambda r + \rho_i \beta_i}{\sqrt{\left( \lambda r \right)^2 + 2 \lambda \rho_i r \beta_i + \rho_i^2}}.
    \label{eq:solution_discrete}
\end{equation}
In the continuum limit, therefore, the solution of the model reads
\begin{eqnarray}
    r &=& \int \frac{\lambda r + \rho \, ( \hat{n} \cdot \hat{\phi} )}{\sqrt{\left( \lambda r \right)^2 + 2 \lambda r \rho \, ( \hat{n} \cdot \hat{\phi} ) + \rho^2}} \, P \left( \rho \right) P \left( \hat{n} \right) d\rho \, d\hat{n} \nonumber \\
    &\equiv& I(r, \hat{\phi}),
    \label{eq:solution}
\end{eqnarray}
a self-consistent equation that, apart from the dependence on conviction, is fully determined by the product $\hat{n} \cdot \hat{\phi}$.

\subsection{Critical point}

The threshold condition for the emergence of a depolarized phase can be obtained by imposing instability for the solution $r = 0$ of Eq.~\eqref{eq:solution}, that is
\begin{equation}
\left.\frac{\partial I(r, \hat{\phi})}{\partial r} \right|_{r = 0} =
\int d\rho \, d\hat{n} \, P(\rho) P(\hat{n}) \frac{\lambda}{\rho} \left[ 1 - (\hat{n} \cdot \hat{\phi})^2 \right] \ge 1,
\end{equation}
The critical point for the onset of instability reads then
\begin{equation}
\lambda_c^\mathrm{MF} = \frac{1}{\int \frac{P \left( \rho \right)}{\rho} P \left( \hat{n} \right) \left[ 1 - (\hat{n} \cdot \hat{\phi})^2 \right] \, d\rho \, d\hat{n}}.
\end{equation}
Interestingly, the dot product can be expressed as $\hat{n} \cdot \hat{\phi} = \cos{\left( \Omega \right)}$, where $\Omega$ is the angle between the unit vectors $\hat{n}$ and $\hat{\phi}$ in the plane containing them. Therefore, the critical point can be written in terms of the cross product $| \hat{n} \times \hat{\phi} | = \sin{\left( \Omega \right)}$, that is
\begin{equation}
\lambda_c^\mathrm{MF} = \frac{1}{\int \frac{P \left( \rho \right)}{\rho} P \left( \hat{n} \right) | \hat{n} \times \hat{\phi} |^2 \, d\rho \, d\hat{n}}.
\label{eq:critical_point}
\end{equation}

\subsection{Critical behavior}

From Eq.~\eqref{eq:D-solution} and setting $P\left( \rho \right) = \delta \left( \rho - 1 \right)$, the self-consistent function for uncorrelated initial opinions with $\hat{n} \cdot \hat{\phi} = \pm 1/\sqrt{D}$ is given by
\begin{equation}
    I \left( r, D \right) = \frac{\lambda r + \frac{1}{\sqrt{D}}}{2\sqrt{\left( \lambda r \right)^2 + 2 \frac{\lambda}{\sqrt{D}}r + 1}} + \frac{\lambda r - \frac{1}{\sqrt{D}}}{2\sqrt{\left( \lambda r \right)^2 - 2 \frac{\lambda}{\sqrt{D}}r + 1}}.
\label{eq:self_uncorrelated}
\end{equation}
Performing a Taylor expansion around $r = 0$ we write
\begin{align}
    I \left( r,D \right) &\simeq \frac{D - 1}{D} \lambda r - \frac{\left( D - 1 \right) \left( D - 5 \right)}{2D^2} \left( \lambda r \right)^3 \nonumber \\
    &+ \frac{3 \left( D - 1 \right) \left( D^2 - 14D + 21 \right)}{8D^3} \left( \lambda r \right)^5 + \mathcal{O} \left( r^7 \right).
\label{eq:Taylor_self}
\end{align}
From Eq.~\eqref{eq:D-threshold}, we find $\lambda_c^\mathrm{MF} = D/\left( D - 1 \right)$. Therefore, the self-consistent equation $r = I \left( r, D \right)$ leads to three different behaviors of the order parameter in the vicinity of the phase transition:
\begin{itemize}
    \item $D < 5$: The third term of the right-hand side of Eq.~\eqref{eq:Taylor_self} can be neglected, leading to
    \begin{eqnarray}
        r \left( \lambda \right) &\simeq& \left[ \frac{2D^2}{\left( D - 1 \right) \left( 5 - D \right)} \frac{1}{\lambda^3} \left( 1 - \frac{\lambda}{\lambda_c^\mathrm{MF}} \right) \right]^{\frac{1}{2}} \nonumber \\
        &=& \frac{1}{\lambda} \sqrt{\frac{2D}{\lambda \left( 5 - D \right)}} \left( \lambda_c^\mathrm{MF} - \lambda \right)^{1/2}.
    \end{eqnarray}
    \item $D = 5$: The second term of the right-hand side of Eq.~\eqref{eq:Taylor_self} vanishes, leading to
    \begin{align}
        r \left( \lambda \right) &\simeq \left[ \frac{8D^3}{3 \left( D - 1 \right) \left( D^2 - 14D + 21 \right)} \frac{1}{\lambda^5} \left( 1 - \frac{\lambda}{\lambda_c^\mathrm{MF}} \right) \right]^\frac{1}{4} \nonumber \\
        &= \frac{1}{\lambda} \sqrt[\leftroot{-1}\uproot{2}\scriptstyle 4]{\frac{8D^2}{3 \lambda \left( 14D - D^2 - 21 \right)}} \left( \lambda - \lambda_c^\mathrm{MF} \right)^{1/4}.
    \end{align}
    \item $D > 5$: The third term of the right-hand side of Eq.~\eqref{eq:Taylor_self} can be neglected, leading to
    \begin{eqnarray}
        r \left( \lambda \right) &\simeq& \left[ \frac{2D^2}{\left( D - 1 \right) \left( D - 5 \right)} \frac{1}{\lambda^3} \left( \frac{\lambda}{\lambda_c^\mathrm{MF}} - 1 \right) \right]^{\frac{1}{2}} \nonumber \\
        &=& \frac{1}{\lambda} \sqrt{\frac{2D}{\lambda \left( D - 5 \right)}} \left( \lambda - \lambda_c^\mathrm{MF} \right)^{1/2}.
    \end{eqnarray}
\end{itemize}
For $D < 5$, therefore, a hysteretic behavior of the order parameter is predicted, typical of explosive phase transitions, while for $D \ge 5$ a continuous transition is expected. Hence, we distinguish a critical dimension $D_c = 5$ for which the depolarization transition goes from being of first-order to second-order.

Interestingly, in the limit $D \to \infty$ the self-consistent function of Eq.~\eqref{eq:self_uncorrelated} becomes
\begin{equation}
    I (r) = \frac{\lambda r}{\sqrt{\left( \lambda r \right)^2 + 1}},
\label{eq:self_correlated}
\end{equation}
which coincides with what is obtained with $\hat{n} \cdot \hat{\phi} = 0$, that is with correlated initial opinions. 
From the self-consistent equation $r = I \left( r \right)$, we find
\begin{equation}
    r (\lambda) = \sqrt{1 - \frac{1}{\lambda^2}},
\label{eq:correlated_solution}
\end{equation}
i.e., $\lambda_c^\mathrm{MF} = 1$ and $\beta = 1/2$.



\section{Heterogeneous mean-field approximation}

The heterogeneous mean-field (HMF) approximation consists in disregarding the actual structure of the network and keeping only a statistical description based on the degree distribution $P(k)$~\cite{pv01a}, giving the probability that a randomly chosen agent has degree $k$ (is connected to $k$ other agents), and the conditional probability $P(k'|k)$ that a random connection from an agent of degree $k$ points to an agent of degree $k'$~\cite{alexei}. 
In the simplest case of uncorrelated networks~\cite{assortative}, where the conditional probability can be written as $P_\mathrm{un}(k'|k) = k' P(k')/\av{k}$~\cite{alexei, dorogovtsev07:_critic_phenom}, the HMF approximation is equivalent to the so-called annealed network approximation~\cite{dorogovtsev07:_critic_phenom}, in which the adjacency matrix can be replaced by its annealed form~\cite{Boguna09}
\begin{equation}
    a_{ij} \simeq \bar{a}_{ij} = \frac{k_i k_j}{N \av{k}}.
\end{equation}
This considers that all network connections are randomly rewired at each microscopic time step of the dynamics, while preserving the same $P(k)$, for any value of $N$. 
Inserting this expression into the model equation Eq.~\eqref{eq:2dim-model}, we obtain
\begin{equation}
    \dot{\theta}_i(t) = \rho_i \sin \left[ \varphi_i - \theta_i(t) \right] + \frac{\lambda k_i}{N \av{k}} \sum_{j = 1}^N k_j \sin \left[ \theta_j(t) - \theta_i(t) \right].
    \label{eq:model_annealed}
\end{equation}

We can introduce now the degree averaged complex global order parameter (see Eq.~\eqref{eq:global_HMF})~\cite{ichinomiya_frequency_2004}
\begin{equation}
r(\lambda) \, \mathrm{e}^{i\psi} = \frac{1}{N \av{k}} \sum_{j=1}^N k_j \mathrm{e}^{i \theta_j(\lambda)},
\label{eq:global_HMF_SM}
\end{equation}
where the mean-field values $r$ and $\psi$ represent the degree of consensus among the population and the average orientation of the agents, respectively. Considering that~\cite{ichinomiya_frequency_2004}
\begin{align}
    \im{r \, \mathrm{e}^{i \left( \psi - \theta_j \right) }} &= r \sin(\psi - \theta_j) \nonumber \\
    &= \im{\frac{1}{N \av{k}} \sum_\ell k_\ell \, \mathrm{e}^{i(\theta_l -\theta_j)}} \nonumber \\
    &= \frac{1}{N \av{k}} \sum_\ell k_\ell \sin(\theta_l -\theta_j),
\end{align}
we can write Eq.~\eqref{eq:model_annealed} as
\begin{equation}
    \dot{\theta}_i = \rho_i \sin(\varphi_i - \theta_i) + \lambda k_i r \sin(\psi - \theta_i),
    \label{eq:model_HMF}
\end{equation}
in which the (apparently independent) orientation variables are coupled by means of the mean-field quantities $r$ and $\psi$. Solving the previous equation for the steady state, we can write the orientation of each agent as a function of $r$ and $\psi$, namely~\cite{ojer_modeling_2023}
\begin{equation}
    \theta_i(r, \psi) = \arctan\left( \frac{ \lambda k_i r \sin(\psi) + \rho_i \sin(\varphi_i) } { \lambda k_i r \cos(\psi) + \rho_i \cos(\varphi_i) }\right).
    \label{eq:theta_HMF}
\end{equation}
Inserting this expression into the degree averaged complex global order parameter Eq.~\eqref{eq:global_HMF_SM}, averaging over the distributions $P(\rho)$ and $P(\varphi)$ and taking the continuous degree approximation, we can write~\cite{ojer_modeling_2023, ichinomiya_frequency_2004}
\begin{eqnarray}
    r  &=& \frac{1}{\av{k}} \int dk \, k P(k) \, d\rho \, P(\rho) \, d\varphi \, P(\varphi) \, \mathrm{e}^{i\left[ \theta(r, \psi) - \psi \right]} \nonumber \\
    &=& \frac{1}{\av{k}} \int dk \, d\rho \, d\varphi \, \frac{ k P(k) P(\rho) P(\varphi) \left[ \lambda k r + \rho \, \mathrm{e}^{i(\varphi - \psi)} \right]}{\sqrt{(\lambda k r)^2 + 2 \lambda k r \rho \cos(\varphi - \psi) + \rho^2}} \nonumber \\
    &\equiv& I_\mathrm{HMF}(\lambda, r, \psi).
\end{eqnarray}
Taking the real and imaginary parts of this expression, we find the relations
\begin{eqnarray}
    r  &=& \re{I_\mathrm{HMF}(\lambda, r, \psi)} \equiv I_r(\lambda, r, \psi), \label{eq:hmf_selfconsistent} \\
    0 &=& \im{I_\mathrm{HMF}(\lambda, r, \psi)} \equiv I_\psi(\lambda, r, \psi).
\end{eqnarray}
The first equation defines the order parameter $r$ self-consistently, while the second relation allows us to compute the value of the average orientation $\psi$.

We can estimate the value of the threshold by looking at the condition for the onset of instability of the solution $r = 0$, namely
\begin{align}
    \left.\frac{\partial I_r}{\partial r} \right|_{r=0} &=
    \int \lambda \, dk \, \frac{k^2 P(k)}{\av{k}} \, d\rho \, \frac{P(\rho)}{\rho} \, d\varphi \, P(\varphi) \sin^2(\varphi - \psi) \nonumber \\
    &\geq 1.
\end{align}
This condition translates into the threshold (see Eq.~\eqref{eq:threshold_HMF})
\begin{equation}
\lambda_c^\mathrm{HMF} = \frac{\av{k}}{\av{k^2}} \lambda_c^\mathrm{MF},
\label{eq:threshold_HMF_SM}
\end{equation}
where
\begin{equation}
    \lambda_c^\mathrm{MF} = \frac{1}{\int_0^\infty d\rho \, \frac{P(\rho)}{\rho} \int_{-\pi}^{\pi} d\varphi \, P(\varphi) \sin^2(\varphi - \psi)}
    \label{eq:threshold_general}
\end{equation}
is the depolarization threshold at the mean-field level~\cite{ojer_modeling_2023}.

\section{Perturbation theory}

For social interactions mediated by a general network, an analytical treatment of the model starts from the definition of a complex local order parameter~\cite{restrepo_onset_2005}
\begin{equation}
    r_j(\lambda) \, \mathrm{e}^{i\psi_j} = \sum_{\ell = 1} ^N a_{j\ell} \, \mathrm{e}^{i \theta_\ell (\lambda)},
    \label{eq:order_local_app}
\end{equation}
with $r_j$ being a local measure of order and $\psi_j$ the local average orientation, computed over the nearest neighbors of an agent. Defining now $\Delta \theta_i \equiv \theta_i - \psi_i$ and $\Delta \varphi_i \equiv \varphi_i - \psi_i$, and omitting the implicit dependence on $\lambda$ for the sake of clarity, we can write the previous equation as
\begin{equation}
    r_j = \sum_{\ell = 1}^N a_{j \ell} \, \mathrm{e}^{i \Delta \theta_\ell} \mathrm{e}^{i \left( \psi_\ell - \psi_j \right)}.
\label{eq:order_perturbation}
\end{equation}
Similarly to Eq.~\eqref{eq:model_HMF}, the model equation can be coupled as
\begin{equation}
    \Delta \dot{\theta}_i = \rho_i \sin(\Delta \varphi_i - \Delta \theta_i) - \lambda r_i \sin(\Delta \theta_i).
\end{equation}
The steady state is then given by
\begin{equation}
    \Delta \theta_i = \arctan{\left( \frac{\rho_i \sin{\left( \Delta \varphi_i \right)}}{\lambda r_i + \rho_i \cos{\left( \Delta \varphi_i \right)}} \right)}.
\label{eq:steady_state}
\end{equation}
Inserting this expression into Eq.~\eqref{eq:order_perturbation}, we obtain~\cite{restrepo_onset_2005}
\begin{align}
r_i &= \sum_{\rho_j} \sum_{\varphi_j} \frac{a_{ij} \cos{\left( \psi_j - \psi_i \right)} \left[ \lambda r_j + \rho_j \cos{\left( \Delta \varphi_j \right)} \right]}{\sqrt{(\lambda r_j)^2 + 2\lambda r_j \rho_j \cos \left( \Delta \varphi_j \right) + \rho^2}} \nonumber \\
&- \sum_{\rho_j} \sum_{\varphi_j} \frac{a_{ij} \sin{\left( \psi_j - \psi_i \right)} \rho_j \sin{\left( \Delta \varphi_j \right)}}{\sqrt{(\lambda r_j)^2 + 2\lambda r_j \rho_j \cos \left( \Delta \varphi_j \right) + \rho^2}}, \label{eq:real_QMF} \\
0 &= \sum_{\rho_j} \sum_{\varphi_j} \frac{a_{ij} \sin{\left( \psi_j - \psi_i \right)} \left[\lambda r_j + \rho_j \cos{\left( \Delta \varphi_j \right)} \right]}{\sqrt{(\lambda r_j)^2 + 2\lambda r_j \rho_j \cos \left( \Delta \varphi_j \right) + \rho^2}} \nonumber \\
&+ \sum_{\rho_j} \sum_{\varphi_j} \frac{a_{ij} \cos{\left( \psi_j - \psi_i \right)} \rho_j \sin{\left( \Delta \varphi_j \right)}}{\sqrt{(\lambda r_j)^2 + 2\lambda r_j \rho_j \cos \left( \Delta \varphi_j \right) + \rho^2}}, \label{eq:imaginary_QMF}
\end{align}
for the real and imaginary parts, respectively. For fully correlated initial opinions with a bimodal orientation distribution given by Eq.~\eqref{eq:angular}, one half of the population holds $\varphi_i = 0$ (state 1), whereas the other half $\varphi_i = \pi$ (state 2). Near the transition to consensus, if the number of connections per agent is large, one half of the neighbors will likely be at state 1 ($\theta_j = 0$) and the other half at state 2 ($\theta_j = \pi$). Therefore, $\psi_i = \pi/2$ $\forall i$ such that $\psi_j - \psi_i = 0$. Eqs.~\eqref{eq:real_QMF} and \eqref{eq:imaginary_QMF} are then reduced to
\begin{align}
r_i &= \sum_{\rho_j} \sum_{\varphi_j} \frac{a_{ij} \left[ \lambda r_j + \rho_j \cos{\left( \Delta \varphi_j \right)} \right]}{\sqrt{(\lambda r_j)^2 + 2\lambda r_j \rho_j \cos \left( \Delta \varphi_j \right) + \rho^2}}, \label{eq:real_QMF_2} \\
0 &= \sum_{\rho_j} \sum_{\varphi_j} \frac{a_{ij} \rho_j \sin{\left( \Delta \varphi_j \right)}}{\sqrt{(\lambda r_j)^2 + 2\lambda r_j \rho_j \cos \left( \Delta \varphi_j \right) + \rho^2}}, \label{eq:imaginary_QMF_2}
\end{align}
respectively. Moreover, for the neighbors at state 1 $\Delta \varphi_j = -\pi/2$, whereas for the ones at state 2 $\Delta \varphi_j = \pi/2$. Hence, Eq.~\eqref{eq:imaginary_QMF_2} for the imaginary part is satisfied. For the real part, approximating the sum of conviction as an integral and summing over all the agents, we obtain
\begin{equation}
r_i = \sum_{j = 1}^N a_{ij} \int d\rho \, P(\rho) \, \frac{\lambda r_j}{\sqrt{(\lambda r_j)^2 + \rho^2}}. \label{eq:real_QMF_3}
\end{equation}
We can determine the threshold value of $\lambda$ by letting $r_i \to 0$. Expanding the right-hand side of the previous equation at first-order, we obtain
\begin{equation}
r_i^{(0)} = \frac{\lambda}{\lambda_c^\mathrm{MF}} \sum_{j = 1}^N a_{ij} r_j^{(0)}, \label{eq:real_expansion}
\end{equation}
where, for a general conviction distribution $P(\rho)$, $\lambda_c^\mathrm{MF}$ is the threshold value Eq.~\eqref{eq:threshold_MF} at the mean-field level. We thus identify the depolarization threshold of the coupling constant with the largest eigenvalue $\Lambda_m$ of the adjacency matrix~\cite{restrepo_onset_2005}, obtaining
\begin{equation}
\lambda_c^\mathrm{P} = \frac{\lambda_c^\mathrm{MF}}{\Lambda_m}.
\label{threshold_QMF_app}
\end{equation}

\section{Numerical results}



Simulations are performed by numerically integrating the model equation Eq.~\eqref{eq:2dim-model} starting from an initial condition $\theta_i(0) = \varphi_i + 10^{-3}\epsilon$, where $\epsilon \in U(-1, 1)$. 
The numerical integration is performed by using the adaptive stepsize algorithm in the function \texttt{odeint} of the Python scientific computing package \texttt{scipy}~\cite{2020SciPy-NMeth}. 
From Eq.~\eqref{eq:scalar_order}, the two-dimensional version of the scalar order parameter can be written as
\begin{equation}
    r(\lambda) = \frac{1}{N} \left| \sum_{j = 1}^N \mathrm{e}^{i\theta_j(\lambda)} \right|.
    \label{eq:order_parameter}
\end{equation}
The order parameter is evaluated at the steady state of the numerical integration, which we define by a difference between consecutive time steps $\Delta r < 10^{-8}$. 
We determined the threshold  $\lambda_c$ as the smallest value of $\lambda$ for which a steady state value $r > 10^{-6}$ is reached.

\section{Empirical networks}

In the following, we briefly describe the different empirical social networks used in our work; 
in Table~\ref{tab:real_networks} we summarize their main topological properties. 
Although some of these networks have a directed nature~\cite{Newman10}, in our simulations we have symmetrized them, rendering them effectively undirected.

\begin{table}[tbp]
    \begin{ruledtabular}
        \begin{tabular}{c|ccccc}
            Network & $N$ & $\av{k}$ & $\chi$ & $\Lambda_m$ & Kind \\
            \hline
            filmtrust & $874$ & $3.00$ & $2.33$ & $14.4$ & Directed \\
            wikitalk1 & $2233$ & $3.47$ & $57.12$ & $35.1$ & Directed \\
            lastFM & $7624$ & $7.29$ & $2.49$ & $38.6$ & Undirected \\
            pgp & $10680$ & $4.55$ & $3.15$ & $42.4$ & Undirected \\
            anybeat & $12645$ & $7.77$ & $49.04$ & $99.9$ & Directed \\
            google & $23628$ & $3.32$ & $112.65$ & $62.6$ & Directed \\
            deezer1 & $28281$ & $6.56$ & $1.47$ & $24.8$ & Undirected \\
            digg & $30398$ & $5.64$ & $3.97$ & $31.2$ & Directed \\
            enron & $36692$ & $10.02$ & $12.98$ & $118.4$ & Undirected \\
            deezer2 & $41773$ & $6.02$ & $0.84$ & $18.6$ & Undirected \\
            slashdot & $51083$ & $4.58$ & $16.80$ & $59.1$ & Directed \\
            brightkite & $58228$ & $7.35$ & $7.66$ & $101.5$ & Undirected \\
            wikitalk2 & $79736$ & $4.60$ & $2307.27$ & $260.8$ & Directed \\
            twitter1 & $102439$ & $6.41$ & $131.14$ & $159.8$ & Directed \\
            douban & $154908$ & $4.22$ & $7.74$ & $42.0$ & Undirected \\
            gowalla & $196591$ & $9.67$ & $30.71$ & $170.9$ & Undirected \\
            emailEU & $265214$ & $2.75$ & $194.00$ & $103.2$ & Directed \\
            wikitalk3 & $338714$ & $2.46$ & $6711.22$ & $285.2$ & Directed \\
            wikitalk4 & $397635$ & $3.14$ & $9642.97$ & $414.5$ & Directed \\
            twitter2 & $465017$ & $3.58$ & $62.19$ & $81.0$ & Directed \\
            youtube1 & $1134890$ & $5.27$ & $92.93$ & $210.4$ & Undirected \\
            hyves & $1402673$ & $3.96$ & $130.82$ & $179.3$ & Undirected \\
            twitter3 & $1896221$ & $3.90$ & $537.53$ & $250.4$ & Directed \\
            flixster & $2523386$ & $6.28$ & $34.08$ & $148.3$ & Undirected \\
            youtube2 & $3223589$ & $5.82$ & $485.75$ & $464.9$ & Undirected \\
        \end{tabular}
    \end{ruledtabular}
    \caption{Topological properties of the real social networks considered: Network size $N$; average degree $\av{k}$; heterogeneity parameter $\chi = \av{k^2}/\av{k}^2 - 1$; largest eigenvalue $\Lambda_m$; kind (directed or undirected).
    }
    \label{tab:real_networks}
\end{table}

\begin{enumerate}
    \item \textbf{filmtrust}~\cite{guo_novel_2013}: user-user trust network of the FilmTrust website, through which each person can recommend movies to their friends.
    \item \textbf{wikitalk1}~\cite{sun_predicting_2016}: communication network of the Welsh Wikipedia website edition, where nodes represent users and edges online messages.
    \item \textbf{lastFM}~\cite{rozemberczki_characteristic_2020}: friendship network of the Last.fm music website, where nodes represent users from Asian countries and edges mutual follower relationships between them.
    \item \textbf{pgp}~\cite{boguna_models_2004}: interaction network of the Pretty Good Privacy encryption algorithm, through which users share confidential information.
    \item \textbf{anybeat}~\cite{fire_link_2013}: friendship network of the AnyBeat online community, where nodes represent users and edges follower relationships between them.
    \item \textbf{google}~\cite{mcauley_learning_2012}: user-user network of the Google+ service, where nodes represent users and edges follower relationships between them.
    \item \textbf{deezer1}~\cite{rozemberczki_characteristic_2020}: friendship network of the Deezer music streaming service, where nodes represent users from European countries and edges mutual follower relationships between them.
    \item \textbf{digg}~\cite{de_choudhury_social_2009}: interaction network of the news website Digg, where nodes represent users and edges replies between them.
    \item \textbf{enron}~\cite{mahoney_community_2009}: communication network of the Enron email dataset, where nodes represent email addresses and edges the emails sent between them.
    \item \textbf{deezer2}~\cite{rozemberczki_gemsec_2020}: friendship network of the Deezer music streaming service, where nodes represent users from Romania and edges mutual follower relationships between them.
    \item \textbf{slashdot}~\cite{gomez_statistical_2008}: interaction network of the technology website Slashdot, where nodes represent users and edges replies between them.
    \item \textbf{brightkite}~\cite{cho_friendship_2011}: friendship network of the Brightkite location-based website, where nodes represent users and edges mutual follower relationships between them.
    \item \textbf{wikitalk2}~\cite{sun_predicting_2016}: communication network of the Catalan Wikipedia website edition, where nodes represent users and edges online messages.
    \item \textbf{twitter1}~\cite{omodei_characterizing_2015}: interaction network of Twitter service during the 2014 People's Climate March, where nodes represent users and edges retweets, mentions and replies between them.
    \item \textbf{douban}~\cite{Zafarani+Liu:2009}: user-user recommendation network of the Douban online website, through which each person can create content with their followers.
    \item \textbf{gowalla}~\cite{cho_friendship_2011}: friendship network of the Gowalla location-based service, where nodes represent users and edges mutual follower relationships between them.
    \item \textbf{emailEU}~\cite{leskovec_graph_2007}: communication network of an European research institution, where nodes represent email addresses and edges the emails sent between them.
    \item \textbf{wikitalk3}~\cite{sun_predicting_2016}: communication network of the Vietnamese Wikipedia website edition, where nodes represent users and edges online messages.
    \item \textbf{wikitalk4}~\cite{sun_predicting_2016}: communication network of the Japanese Wikipedia website edition, where nodes represent users and edges online messages.
    \item \textbf{twitter2}~\cite{choudhury_how_2010}: friendship network of Twitter service, where nodes represent users and edges follower relationships between them.
    \item \textbf{youtube1}~\cite{yang_defining_2015}: friendship network of the YouTube video-sharing website, where nodes represent users and edges mutual follower relationships between them.
    \item \textbf{hyves}~\cite{Zafarani+Liu:2009}: friendship network of the Hyves website, where nodes represent users and edges mutual follower relationships between them.
    \item \textbf{twitter3}~\cite{de_domenico_unraveling_2020}: interaction network of Twitter service during the 2015 Paris attacks, where nodes represent users and edges retweets, mentions and replies between them.
    \item \textbf{flixster}~\cite{Zafarani+Liu:2009}: friendship network of the Flixster movie rating site, where nodes represent users and edges mutual follower relationships between them.
    \item \textbf{youtube2}~\cite{mislove_measurement_2007}: friendship network of the YouTube video-sharing website, where nodes represent users and edges mutual follower relationships between them.
\end{enumerate}


%

\end{document}